\documentclass[aip,jap,reprint,superscriptaddress]{revtex4-1}
\usepackage{amsmath}
\usepackage{graphicx}
\usepackage{dcolumn}
\usepackage{bm}
\usepackage{natbib}

\begin{document}

\title{Analysis of laser shock experiments on precompressed samples using a quartz reference and application to warm dense hydrogen and helium}

\author{Stephanie Brygoo}
\email{stephanie.brygoo@cea.fr}
\affiliation{CEA, DAM, DIF, F-91297 Arpajon, France}

\author{Marius Millot}
\email{millot1@llnl.gov}
\affiliation{Lawrence Livermore National Laboratory, Livermore, California 94550, USA}
\affiliation{University of California-Berkeley, Berkeley, California 94720, USA}

\author{Paul Loubeyre}
\affiliation{CEA, DAM, DIF, F-91297 Arpajon, France}

\author{Amy E. Lazicki}
\author{Sebastien Hamel}
\author{Tingting Qi}
\author{Peter M. Celliers}
\author{Federica Coppari}
\author{Jon H.Eggert}
\author{Dayne E. Fratanduono}
\affiliation{Lawrence Livermore National Laboratory, Livermore, California 94550, USA}

\author{Damien G. Hicks}
\affiliation{Centre for Micro-Photonics, Swinburne University of Technology, Hawthorn, VIC 3122, Australia}

\author{J. Ryan Rygg}
\author{Raymond F. Smith}
\author{Damian C. Swift}
\author{Gilbert W. Collins}
\affiliation{Lawrence Livermore National Laboratory, Livermore, California 94550, USA}
\author{Raymond Jeanloz}
\affiliation{University of California-Berkeley, Berkeley, California 94720, USA}

\begin{abstract} 
Megabar (1 Mbar = 100 GPa) laser shocks on precompressed samples allow reaching unprecedented high densities and moderately high $\sim 10^3-10^4$K temperatures. We describe here a complete analysis framework  for the velocimetry (VISAR) and pyrometry (SOP) data produced in these experiments. Since the precompression increases the initial density of both the sample of interest and the quartz reference for pressure-density, reflectivity and temperature measurements, we describe analytical corrections based on available experimental data on warm dense silica and density-functional-theory based molecular dynamics computer simulations. Using our improved analysis framework we report a re-analysis of previously published data on warm dense hydrogen and helium, compare the newly inferred pressure, density and temperature data with most advanced equation of state models and provide updated reflectivity values.
\end{abstract}

\maketitle

\pagestyle{empty}

\section{Introduction}

There is a great interest for measuring the properties of warm dense low-Z molecular systems, motivated by planetary implications and the fundamental understanding of the warm dense matter regime: pressures of a few hundreds of GPa (= a few Mbar) and temperatures $\sim$ 10$^4$ K$\sim$1eV. Complex structural and chemical modifications from the molecular fluids to the warm dense plasma, are expected in this domain such as the existence of a first order transition between the molecular fluid and the plasma state in dense hydrogen (known as the plasma phase transition) or a superionic state in dense H$_{2}$O. Since 2003 a new approach combining static and dynamic compression techniques by launching strong shockwaves in precompressed samples\cite{Loubeyre03,Jeanloz07} has been developed and allows to explore those new extreme conditions of matter. As a few GPa precompression can induce significant density increases in compressible fluids, the locus of shock states (Hugoniot) accessible by the subsequent shock compression reaches lower temperatures and higher densities, as demonstrated on hydrogen and helium\cite{Eggert08,loubeyre, Celliers2010}.

\begin{figure}[t]
	\centerline{\includegraphics[width=8cm,trim=3 2 1 0,clip]{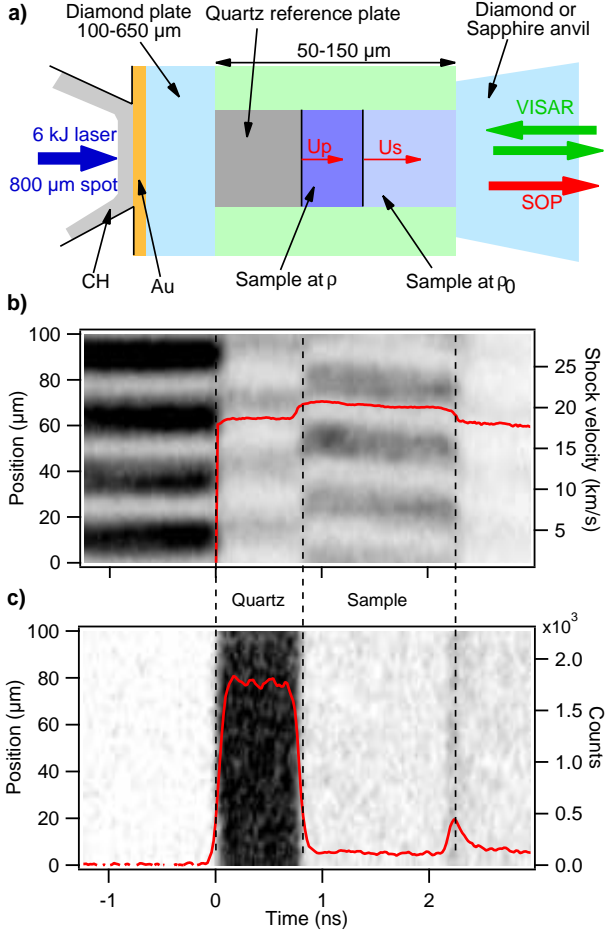}}
	\caption{Experimental setup and raw velocimetry and pyrometry data measured at the Omega laser (LLE, Rochester, NY). a) Sketch of the diamond anvil cell (DAC). Direct drive laser ablation launches a shock in the precompressed target assembly and its propagation is monitored with velocimetry (VISAR) and pyrometry (SOP) through the back anvil. Typical pulses shapes for the laser are 1 ns with 1-6 kJ. Raw velocimetry (b) and pyrometry (c) images with superimposed shock velocity and counts (right scales) are presented. For clarity, the arrival of the shock in the quartz is chosen as the origin of the time scale. Depending on the anvil thickness, this event occurs $\sim 5-25$ ns after the drive laser pulse. }
	\label{fig:Image1}
\end{figure}

A typical configuration is sketched in Fig. \ref{fig:Image1}: a sample is precompressed in a diamond anvil cell before being submitted to a strong shock compression generated by direct laser ablation of a thin plastic polymer layer deposited on one of the anvils\cite{Loubeyre03,Jeanloz07}.
The shock wave propagation is monitored with ultrafast Doppler velocimetry\cite{Celliers04} (VISAR) and pyrometry \cite{Miller07,Millot2015a} (SOP) through the back anvil. We use a quartz plate precompressed with the sample as an \textit{in-situ} reference for the impedance-matching procedure that allows obtaining pressure-density equation of state data from the velocimetry measurements. We also use the reflectivity and emission from the shock front during its transit in the quartz as a reference for the reflectivity and temperature measurements\cite{Millot2015a,reason}. Important progress have been made recently in the characterization of quartz under shock compression\cite{KnudsonRelease,Millot2015a,Qi2015} and its release from shocked states. However, in the case of precompressed targets, the quartz is not following the principal Hugoniot and the higher initial density needs to be accounted for. 

In the following, relying on a better understanding of shocked compressed SiO$_2$\cite{KnudsonRelease,Millot2015a,Qi2015}, we describe an improved analysis framework where corrections to the principal Hugoniot are presented and we show a re-analysis of previously published data on hydrogen\cite{Loubeyre03,loubeyre} and helium\cite{Eggert08,Celliers2010} to document the changes in the inferred data according to this new framework. An appendix contains details on the characterization of the initial state and the associated uncertainties.

\section{Analysis framework}
\subsection{Pressure-density equation of state measurements}
\subsubsection{VISAR velocimetry of strong reflecting shocks }   

We use a line-imaging streaked velocity interferometer system for any reflector (VISAR): an interferometric technique which records a phase shift proportional to the velocity of fast moving reflectors\cite{Barker65,Barker70,Barker72}. When strong enough shock waves propagate in transparent media such as oxides or low-Z compounds they can produce a reflecting shock front. In this case, VISAR offers a line-imaging time-resolved record of the shock speed with better than 1\% accuracy\cite{Celliers04}. We can then obtain the shock velocities just before and after the shock crosses the interface between the quartz reference and the sample: U$_S^Q$ and U$_S^S$.

Note that the true velocity of the shock front is in fact the ratio of the apparent velocity inferred from the fringe shift to the refractive index of the medium at rest\cite{Celliers04}. The knowledge of the refractive index of the precompressed sample and quartz reference plate is therefore required to obtain the shock velocity in the quartz U$_S^Q$(t) and in the sample U$_S^S$(t), and the uncertainty needs to be propagated. The formulas used for quartz are indicated in the appendix.

\subsubsection{Equation of state determination by shock impedance matching } 

\begin{figure}[bp]
	\centerline{\includegraphics[width=8cm,clip]{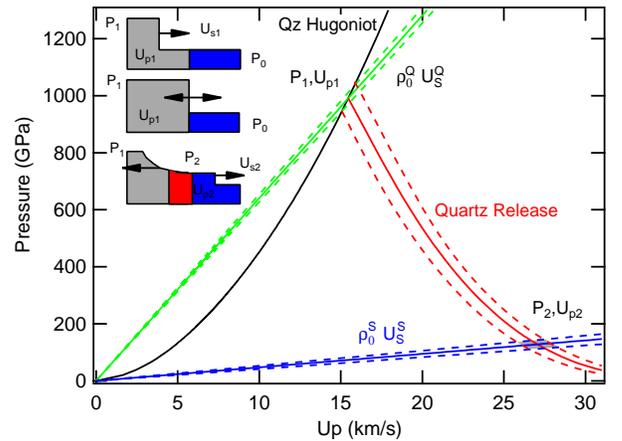}}
	\caption{Impedance matching construction: shock Hugoniot of the quartz standard (black), release isentrope (red) and Rayleigh lines having a $\rho_0$U$_S$ slope in the (P,u$_P$) plane (green and blue). The uncertainty in determining the quartz velocity U$_S^Q$ (green dashed lines) gives a set of different possible first shock states (P$_1$, u$_{p1}$) from which we calculate different possible release curves (red) which will intersect the possible Rayleigh lines for the sample (blue lines) yielding an area of possible final states for the sample (P$_2$,u$_{p2}$). Inset: Sketch of the shock wave interaction with the quartz (grey) - sample (blue) interface.}
	\label{fig:IM}
\end{figure}

When the shock wave encounters the quartz-sample interface, it is partly transmitted due to the shock impedance mismatch and a release wave (or a reshock depending on the impedance of each sample) is sent propagating backward into the quartz standard to ensure the continuity of the pressure and particle velocity at the interface. A graphical construction (Fig.~\ref{fig:IM}) illustrates the derivation of the transmitted pressure and particle velocity. Knowing the Hugoniot of the quartz standard and using the Rankine-Hugoniot conservation relations:
\begin{align}
\rho&=\rho_0 \frac{U_S}{U_S- u_p}\\ 
P&=P_0+\rho_0 U_S u_p\\
e&=e_0+\frac{1}{2}(P+P_0)(1/\rho_0-1/\rho)
\end{align} 

the measurement of U$_S^Q$ determines the incident shock state (P$_1$, u$_{p1}$). The unknown transmitted shock state in the sample is therefore the intersection of the isentropic release path of the quartz standard from (P$_1$, u$_{p1}$) and the Rayleigh line P$_2$=P$_0$+$\rho_0^S$U$_S{^S}$u$_p$ where P$_0$ and U$_S^S$ have been measured and $\rho_0^S$ is inferred from P$_0$ using the sample's equation of state determined at ambient temperature.

\subsubsection{Available experimental shock data and precompression correction model}

\begin{figure}[tp]
	\centerline{\includegraphics[width=8cm,trim=0 0 0 0,clip]{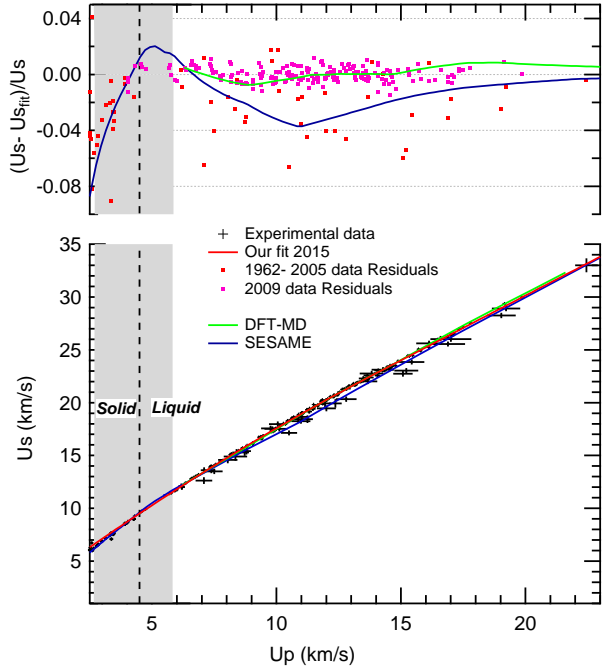}}
	\caption{Quartz principal Hugoniot: shock velocity U$_S$ versus particle velocity u$_p$. Bottom: experimental data\cite{Adadurov62,Altshuler65,Trunin71,Pavlovskii76,VanThiel77, Marsh80,Trunin94,Hicks05,Knudson} (black crosses), fit to the data (red) and calculated Hugoniot from Sesame (dark blue) and DFT-MD (green). Top: Fit residuals (U$_S$-U$_{S~fit}$)/U$_S$ as a function of u$_p$. Red dots corresponds to data until 2005\cite{Adadurov62,Altshuler65,Trunin71,Pavlovskii76,VanThiel77, Marsh80,Trunin94,Hicks05,Knudson} and pink dots correspond to the most recent ones\cite{Knudson}. Limit between solid and liquid shocked silica is represented by the vertical dashed line. Gray area: quartz reflectivity is below 2\% making direct shock velocity measurement with VISAR challenging.}
	\label{fig:UsUp}
\end{figure}

The pressure-density relationship of shock compressed quartz in the high pressure fluid regime has been well characterized\cite{Adadurov62, Altshuler65,Trunin71,Pavlovskii76,VanThiel77, Marsh80, Trunin94,Hicks05,Knudson}. We use a weighted, piecewise polynomial U$_S$-u$_p$ fit of all existing data in the liquid domain\cite{Adadurov62, Altshuler65,Trunin71,Pavlovskii76,VanThiel77, Marsh80,Trunin94,Hicks05,Knudson}: see Table \ref{tab:summary} and
Figure~\ref{fig:UsUp}. The difference with the published fit\cite{Knudson}, obtained using only Z-pinch data, is less than 1\%.

Calculated Hugoniot obtained with an analytical equation of state model\cite{Kerley99} (Sesame) and recent density functional theory based molecular dynamics (DFT-MD) simulations using the AM05 exchange-correlation functional are also presented\cite{Qi2015}. As previously shown, the AM05 DFT-MD simulations capture quite well the pressure-density shock compressibility of warm dense SiO$_2$ along the quartz and fused silica Hugoniot\cite{Qi2015}.

\begin{figure}[tp]
	\centerline{\includegraphics[width=7.9cm,trim=0 0 0 0,clip]{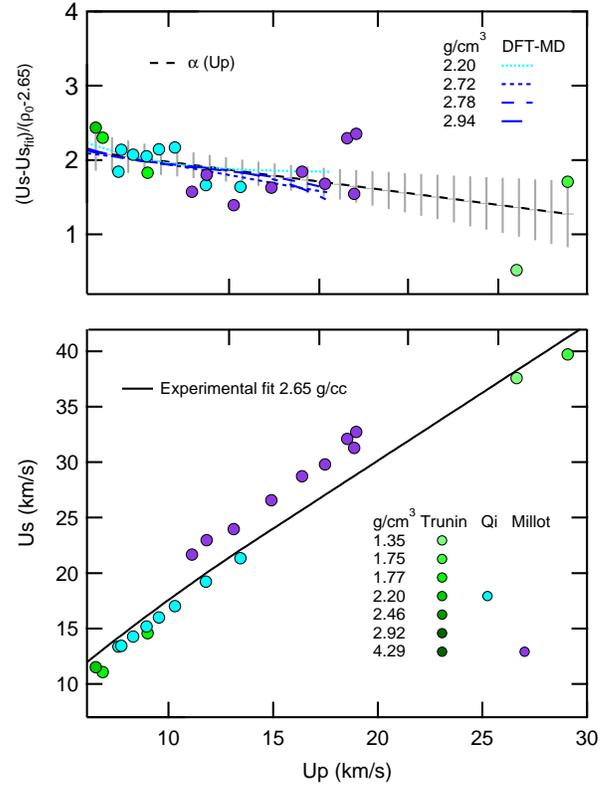}}
\caption{Precompression correction on the quartz U$_S$-u$_p$ Hugoniot. Bottom: shock data for different silica starting materials - porous silica\cite{Trunin70, Trunin71, Trunin94, Trunin94b} (green), fused silica\cite{Qi2015} (cyan), and  stishovite\cite{Millot2015a} (purple) - and fit to the quartz Hugoniot. Top: density scaled deviation from the quartz Hugoniot defined as (U$_S$-U$_{S~fit~2.65}$)/($\rho_{0}$-2.65) for the data shown in the bottom panel and our model (dashed black line). Initial densities of 2.72, 2.78 and 2.94 g/cm$^3$ correspond to 1, 2 and 5 GPa precompressions, fused silica density is 2.20 g/cm$^3$. Sesame model gives similar results to the DFT-MD (not shown).}
\label{fig:UsUpAlpha}
\end{figure}

To describe the Hugoniot of the precompressed reference we use the experimental data available at standard density 2.65 g/cm$^{3}$ and apply a small correction. It has been observed that for shock pressures above 150 GPa, the U$_S$-u$_p$  relationship for various allotropic forms and various porosities of SiO$_{2}$ (initial density ranging from 1.15 g/cm$^{3}$ to 4.31 g/cm$^{3}$) can be approximated by a set of parallel lines shifted by an offset that depends linearly on the initial density\cite{Trunin70, Trunin71, Trunin94, Trunin94b}. In fact, recent high-precision measurements on fused silica\cite{Qi2015} ($\rho_0$=2.20 g/cm$^3$) and stishovite\cite{Millot2015a} ($\rho_0$=4.29 g/cm$^3$) shocked to the dense fluid state, as well as the Sesame and DFT-MD models (see Fig.~\ref{fig:UsUpAlpha}) suggest that a slightly more complex correction can describe these data more accurately. 

Using the available experimental data for the different silica starting materials, we derive a correction to estimate U$_S$(u$_p$,$\rho_0$) as a correction from the quartz Hugoniot U$_S$(u$_p$,$\rho_0$=2.65)  (all velocities in km/s and densities in g/cm$^3$):
\begin{equation}
U_S(u_p,\rho_0) = U_S(u_p,\rho_0=2.65) + \alpha(u_p)(\rho_{0}-2.65)
\end{equation}
with
\begin{equation}
\alpha(u_p) = 2.3(\pm0.4)- 0.037 (\pm0.027)~u_p
\end{equation}
Only data in the liquid phase were used for the fit. The density change of the quartz reference (having a bulk modulus of 37 GPa) is however limited: it only amounts for ~10 \% at a challenging precompression of 5 GPa. 

\begin{figure}[tp]
	\centerline{\includegraphics[width=8cm,trim=0 0 0 0,clip]{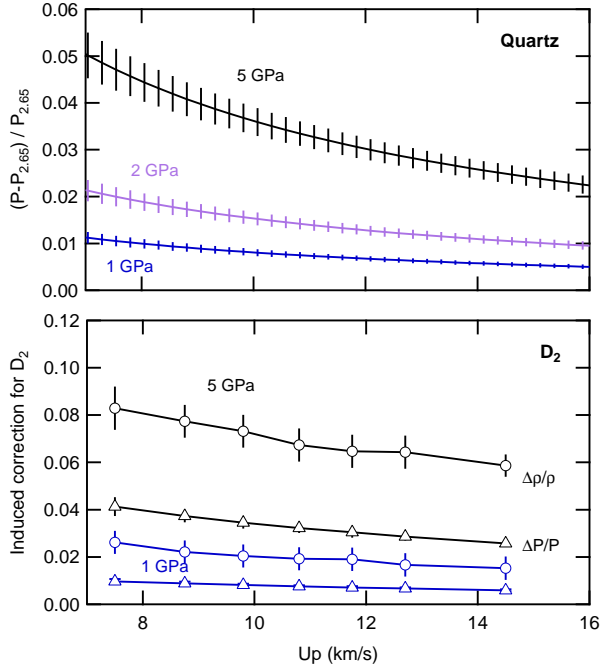}}
	\caption{Magnitude of the precompression correction on the quartz U$_S$-u$_p$ Hugoniot. (Top) Effect on the inferred shock pressure in the quartz reference (P($\rho_{0}$)-P(2.65))/P(2.65). (Bottom) Effect on the inferred final pressure (triangles) and density (circles) for D$_2$ precompressed to 1 GPa (blue) and 5 GPa (black). The error bars reflect the systematic uncertainties arising from the uncertainty in determining $\alpha(u_p)$}
	\label{fig:CorrectionAlpha}
\end{figure}

We show the magnitude of the correction obtained with our model in Fig.~\ref{fig:CorrectionAlpha} (Top). For an initial density of 2.72 g/cm$^3$ (P$_{0}$=1 GPa), the relative difference in shock pressure in the quartz (P($\rho_{0}$)-P(2.65))/P(2.65) is less than 2\%, increasing to a few \% with  an initial density of  2.94 g/cm$^3$ corresponding to a 5 GPa precompression. 

Note that, due to the large impedance mismatch between the quartz and typical samples, larger changes are observed in the inferred quantities from the impedance matching procedure. In Fig. ~\ref{fig:CorrectionAlpha} (Bottom) we present the magnitude of the change in pressure and density defined as: 
\begin{equation}
\frac{P^{D2}_{\rho_{qz}=\rho_{0}}-P^{D2}_{\rho_{qz}=2.65}}{P^{D2}_{\rho_{qz}=\rho_{0}}}
\end{equation}
and
\begin{equation}
\frac{\rho^{D2}_{\rho_{qz}=\rho_{0}}-\rho^{D2}_{\rho_{qz}=2.65}}{\rho^{D2}_{\rho_{qz}=\rho_{0}}}
\end{equation} 
for a deuterium sample at two different precompressions P$_{0}$=1 GPa and 5 GPa (we used Caillabet et al.\cite{Caillabet} equation of state for D$_2$). The correction of 4\% in quartz velocity thus corresponds to a correction of 8\% in shock density for deuterium precompressed to 5 GPa.

\subsubsection{Quartz release model}
Once the shock state in the quartz reference P$_1$ has been determined from the measurement of the shock velocity and the initial pressure, one has to determine the possible final states for the reference by computing the release curve from P$_1$. In contrast with the great wealth of data on the principal Hugoniot, there are very few release measurements available for the quartz principal Hugoniot, and none for higher density polymorphs. 

Previous studies of laser shocks on precompressed samples\cite{Eggert08, loubeyre} estimated the release by calculated the reflected Hugoniot and applying a correction depending on a constant Gruneisen\cite{gruneisen,Barrios2010,Hicks06} $\Gamma\simeq0.64$ or $0.66$.

An improved release model has been recently developed based on DFT-MD simulations and shock experiments\cite{KnudsonRelease}. The release isentrope is parameterized as a Mie-Gruneisen correction from an effective reflected Hugoniot, using an effective Gruneisen parameter $\Gamma_{eff}$. The reflected Hugoniot is defined by a linear U$_S$-u$_p$ relationship: U$_S$=c$_{1}$+s$_1$u$_p$ which makes the derivation of the release isentrope analytical. The slope s$_1$ is fixed to 1.197 and c$_1$ can therefore be uniquely determined from P$_1$. The release isentrope can then be obtained using the parametrization of $\Gamma_{eff}$ provided in Ref. \cite{KnudsonRelease}. Note that, in the precompression case, the derivation of c$_{1}$ needs to account for the initial precompression P$_0>$0:
\begin{equation}
	c_{1}=\frac{P_{1}-P_{0}}{\rho_{0}^Qu_{p1}}-s_1u_{p1}
\end{equation}
and $P_{1}$ and $u_{p1}$ are determined using the precompressed hugoniot.

\begin{figure}[tp]
	\centerline{\includegraphics[width=8cm,trim=0 0 0 0,clip]{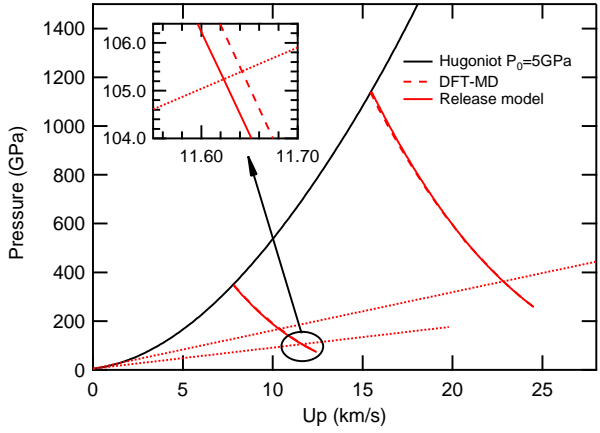}}
	\caption{Release model for precompressed Hugoniot. Example of the impedance matching construction in the P-u$_p$ plane for a 5 GPa initial pressure: precompressed quartz Hugoniot (solid black), release isentropes from 348 GPa (Us$_{Qz}$=15km/s) and 1139 GPa (Us$_{Qz}$=25km/s) obtained from the DFT-MD (dashed red) and the release model from Ref. \cite{KnudsonRelease}(solid red) modified to take into account the initial pressure and Rayleigh lines for the two corresponding shocks in liquid deuterium (dotted red). Inferred u$_p$ for the deuterium is determined at the intersection of the Rayleigh line and the release curves.}
	\label{fig:DiffUp}
\end{figure}

We tested the validity of using this model (which had only been tested\cite{KnudsonRelease} when P$_0$=0) to compute the release states of quartz shocked from a precompressed state by comparing the calculated release paths with isentropic releases computed from DFT-MD simulations. Note that the DFT-MD simulations were found to be in good agreement with the experiments in building the release model in Ref. \cite{KnudsonRelease}. Fig.~\ref{fig:DiffUp} presents the difference in inferred u$_p$ along the isentrope for D$_2$ at a 5 GPa precompression for two shocks 15 km/s and 25 km/s in quartz. The difference is smaller than 1\% independantly of the initial pressure or the shock along most of the isentrope. The overall difference being smaller than the numerical noise introduced by interpolated the DFT-MD pressure and energy, we conclude that this model seems appropriate to describe release states from precompressed shock states. In the future, improved equation of state might confirm this assumption or provide a more accurate way of obtaining the quartz release paths.

\subsubsection{Re-analysis of helium, deuterium and hydrogen pressure-density data}

\begin{figure}[htp]
	\includegraphics[width=8cm,trim=0 0 0 0,clip]{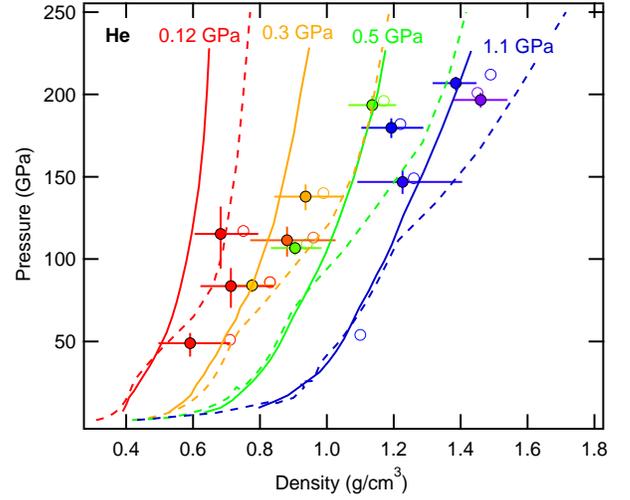}
	\caption{Shock pressure versus density data for warm dense helium: open and solid symbols are respectively the published data\cite{Eggert08} and the re-analyzed one. The solid and dashed lines are DFT-MD calculations\cite{Militzer} from Militzer \textit{et al.} and the chemical SCVH model\cite{SCVH}. Red, orange, green and blue indicate the initial pressure of 0.12 GPa (0.123 g/cm$^3$), 0.30 GPa ( 0.225 g/cm$^3$), 0.50 GPa ( 0.296 g/cm$^3$) and 1.10 GPa (0.412 g/cm$^3$) respectively. } \label{fig:figHeb}
\end{figure}

We present in Figure~\ref{fig:figHeb} the shock equation of state data on warm dense helium\cite{Eggert08} first reported in 2008. Three contributions account for the difference between the values originally reported\cite{Eggert08} and the new ones. First, the function $\alpha$ was set to a constant value of 2.42 instead of a varying function with u$_p$. The precompression being below 1.25 GPa, the difference is below 0.35\% which is negligible. As a comparison, for a precompression of 5 GPa, the difference would be aroung 1\%. Using the new fit for the Hugoniot has a larger effect on the final density which, on average, \textit{decreases} by 10\% and by 4\% in pressure. In contrast, the new release model has the opposite effect contributing to an \textit{increase} in density of around 6\%. So overall only a 4\% \textit{decrease} is density is observed. With the improved analysis, at low precompressions, the experimental data are in better agreement with both DFT-MD calculations\cite{Militzer} and the chemical model SCVH\cite{SCVH}, but at higher precompressions DFT-MD calculations reproduce better the experimental data. 

\begin{figure}[htp]
\includegraphics[width=80mm,trim=0 0 0 0,clip]{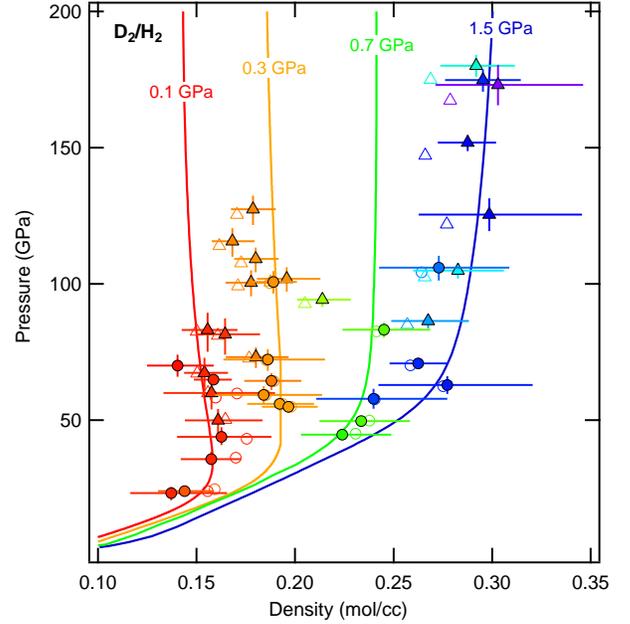}
\caption{Shock pressure versus molar density for warm dense hydrogen and deuterium: open and solid symbols are respectively the published data\cite{loubeyre} and the re-analyzed one. Triangles are D$_{2}$ data and circles H$_{2}$ data. The solid lines are DFT-MD calculations from Caillabet \textit{et al.}\cite{Caillabet}. Red, orange, green and blue indicate the initial pressure of 0.13 GPa (0.029 mol/cm$^3$), 0.30 GPa (0.044 mol/cm$^3$), 0.70 GPa (0.061 mol/cm$^3$) and 1.50 GPa (0.079 mol/cm$^3$) respectively.} \label{fig:figD2H2}
\end{figure}

We present in Figure~\ref{fig:figD2H2} the shock equation of state data on warm dense hydrogen and deuterium\cite{loubeyre} first reported in 2012. 
The difference between the previously reported values and the new ones comes mainly from the new release model. Depending on the shock impedance of the sample, the final density is either increased or decreased by a few \% relative to the initial report. The agreement between the data and the latest DFT-MD calculations on hydrogen isotopes\cite{Caillabet} is improved. This seems to lift the small systematic discrepancy between the simulations and the data.

\subsubsection{Evolution of the random and systematic uncertainties}
Random uncertainties coming from technical limitations in the experimental accuracy of the observables - shock velocities, initial pressure - are sometimes important but are generally getting smaller as improved diagnostics are used and better experimental procedures are developed. 

Systematic uncertainties arise from the models used to determine the initial density and refractive index of both the quartz and the sample, as well as describing the behavior of the quartz under shock and release (including the precompression corrections).

The uncertainty in the initial density  $\rho_{0}$  is a combination of the uncertainty in measuring the initial pressure P$_0$ by ruby luminescence ($\Delta$ P=0.03 GPa independently of the initial pressure and in the absence of pressure gradients within the precompressed sample) and the uncertainty in the static compression equation of states. Similar reasoning apply to the refractive index. Since the main contribution comes from the constant uncertainty in the initial pressure, as the precompression is increased from 0.1 Gpa to a few GPa, the importance of these uncertainty sources strongly decreases.

The reduction of the uncertainty on quartz shock and release behavior is the most important progress of the model described in this work compared to the previous analysis of early data on hydrogen and helium. In addition, the improved correction for the higher initial density contributes to more accurate data. For example, for hydrogen, on average, the error in pressure drops from 7\% to 5\% and in compression, from 12\% to 10\%. All re-analyzed data are presented in Tables \ref{tab:He}, \ref{tab:H2} and \ref{tab:D2}.

\subsection{Temperature of shocked compressed SiO$_2$}
\subsubsection{Diagnostics and data analysis}

When monitoring reflecting shock fronts, the streaked optical pyrometer (SOP) images the thermal emission of the propagating shock front over a small spectral range\cite{Miller07,Millot2015a}, first in the quartz and then in the sample (Fig. \ref{fig:Image1}). To determine the temperature from the measured thermal emission, we assume a grey-body approximation for the spectral radiance, I=A$\epsilon$($\lambda$) [e$^{hc /\lambda kT}$-1 ]$^{-1}$ where $\epsilon$($\lambda$)=$\epsilon$ is the emissivity given by (1-R), R the measured optical reflectivity, A is a system calibration constant that incorporates the transfer function of the optical system and the response of the detector. 

Inverting this expression to solve for temperature gives, T =T$_{0}$ [ln($\epsilon$($\lambda)$ A /I +1)]$^{-1}$, where T$_{0}$ = hc/ $\lambda_{0}$kT is a calibration parameter related to the wavelength of the spectrometer peak sensitivity (T$_{0}$  $\sim$ 1.9 eV at $\lambda_{0}$= 650 nm). Since the temperature determination is made relative to the quartz reference, the temperature in the sample can be determined from the ratio of the signal levels observed in the quartz and the sample, such that the system calibration constant A drops out of the expression,
 T$_{S}$ =T$_{0}$ [ln(e$^{T_{0}/T_{Q}}$-1)(I$^{*}_{Q}$/I$^{*}_{S}$+1)]$^{-1}$ 
with  I$^{*}$=I$_{ADU}$/(1-R),  I$_{ADU}$ being the analog-to-digital counts associated with the observed signal, and R being the reflectivity measured with the VISAR. 

Knowing the shock velocity in quartz, the temperature in quartz, T$_{Q}$, has to be determined from the calibrated function T$_{Q}$(Us). Then, one can obtain the 
temperature T$_{S}$ in the sample.

\subsubsection{Available experimental shock data and precompression correction model}

\begin{figure}[tp]
	\centerline{\includegraphics[width=8cm,trim=0 0 0 0,clip]{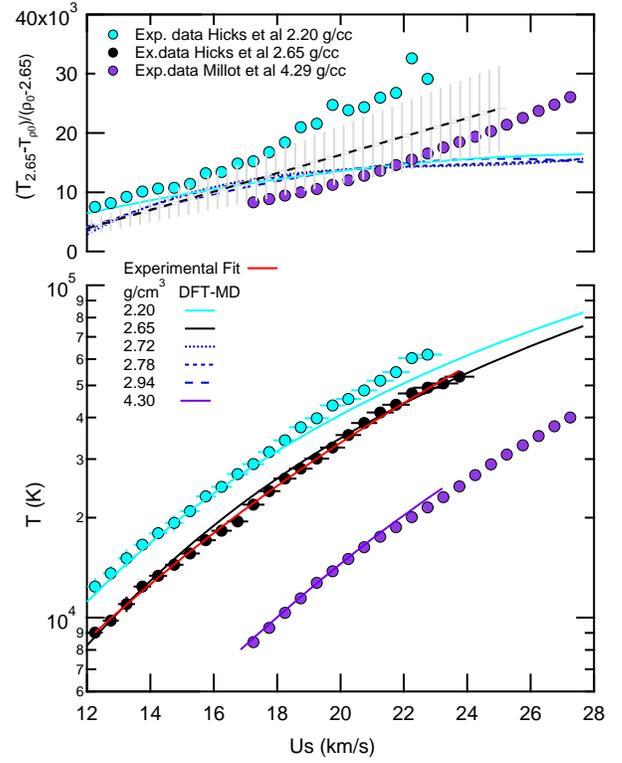}}
\caption{Shock temperature as a function of the shock velocity for warm dense SiO$_2$. Bottom: Experimental data for quartz\cite{Hicks06,Millot2015a}, fused silica\cite{Hicks06} and stishovite\cite{Millot2015a} and DFT-MD (this work). Top: Precompression model and density scaled data relative to the quartz Hugoniot (dashed black with grey error bars). } \label{fig:TvsUs}
\end{figure}

The temperature along the principal Hugoniot T$_{Q}$(U$_S$) has been measured by Hicks et al.\cite{Hicks06} (black circles in Figure~\ref{fig:TvsUs}) and is well represented between 12 and 23 km/s by a power law:
\begin{equation}
T(2.65, Us)(K) =1860 + 3.56 Us^{3.036}
\end{equation}

Due to its higher initial density\cite{Jeanloz07}, the shock temperature of precompressed quartz will be lower at a given shock pressure. We propose a simple parametrization of the shock temperature for precompressed quartz based on experimental data on fused silica (2.20 g/cm$^3$) and stishovite (4.29 g/cm$^3$) (light blue circles and light purple circles in Figure~\ref{fig:TvsUs}). The DFT-MD is found to capture well the experimental data for the three starting densities. So does the Sesame EOS\cite{Kerley99} at high pressure, once silica is considered fluid and dissociated. As it was done for the principal Hugoniot in the pressure-density plane, several Hugoniots at different densities are calculated and compared to estimate the shift in temperature expected with a higher initial density. The Sesame model and the DFT-MD simulations give very similar results and trends. We observe that a simple density scaling allows us to describe the difference in shock temperature between the quartz Hugoniot and either the fused silica or the stishovite Hugoniot (black curve in Figure~\ref{fig:TvsUs}):
\begin{equation}
	T(\rho_{0},U_S)=T(2.65,U_S)-(-14786 + 1555 U_S)(\rho_{0}-2.65)
\end{equation}
Given the reduced and sparse set of data a relative uncertainty on this correction $\sim \pm30\%$ seems reasonable: this allows the model to describe relatively well the experimental data for lower and much higher initial densities as well as the DFT-MD results.

\subsection{Reflectivity of shocked compressed SiO$_2$}
\subsubsection{Diagnostics and data analysis}

In addition to extracting the shock velocity from VISAR fringe pattern shifts we can also measure the reflectivity of the moving reflecting interface from the intensity of the fringes. In the case of a reflecting shock, the reflectivity (at the wavelength of the probe laser) is due to a mismatch of complex refractive index between the shock compressed material (index n) and the precompressed material (index n$_0$) :
\begin{align}
R=\frac{|n-n_0|^2}{|n+n_0|^2}
\end{align}
A relative measurement of the reflectivity of the shock front in the sample compared to the shock front reflectivity in the quartz reference can be obtained easily from the ratio of intensities of the VISAR fringes in quartz and in the sample. When the shock front is in the quartz, the measured intensity of the VISAR fringes I$_{Q}$ is given by: I$_{Q}$=I$_{P}$R$_{Q}$(U$_{S}$)f$_{T}$
where I$_{P}$ is the incident probe laser intensity at the shock front, R$_{Q}$(U$_{S}$) is the calibrated reflectivity of the shock front in quartz and f$_{T}$ is the unknown transfer function of the optical system. When the shock is in the sample, the intensity of the VISAR fringes is given by: $ I_{S}= I_{P}R_{S}f_{T}$ where R$_{S}$ is the reflectivity of the sample shock front.  Combining these two equations gives: R$_{S}$=R$_{Q}$(U$_{S}$)I$_{S}$/I$_{Q}$.
A reflection at the quartz/sample interface can come from an index mismatch of the precompressed states and can be easily modeled knowing the influence of the precompression on the respective refractive index. It is usually less than 2\% and it has been in this case neglected.

This relative measurement allows for accurate characterization of the optical reflectivity of the shocked sample even in the presence of strong variations of the transparency of the back diamond anvil/window that can be caused by the interaction of the high-energy drive-laser with the diamond anvil cell target and the ablation plasma. The measured reflectivity can then be used to estimate the temperature using the grey body approximation.

\subsubsection{Available experimental shock data and precompression correction model}

\begin{figure}[tp]
	\centerline{\includegraphics[width=8cm,trim=0 0 0 0,clip]{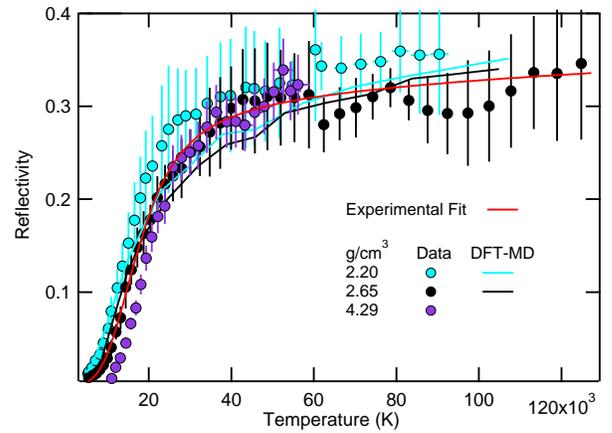}}
	\caption{Shock reflectivity as a function of temperature: black, light gray and dark gray correspond respectively to experimental data on quartz\cite{Hicks06}, fused silica\cite{Hicks06} and stishovite\cite{Millot2015a}. The solid dark blue and light blue are reflectivity obtained from the DFT-MD simulations for quartz and fused silica.} \label{fig:RvsT}
\end{figure}

We show on Figure~\ref{fig:RvsT} experimental data for shock reflectivity along three different Hugoniot starting with fused silica\cite{Hicks06}, quartz\cite{Hicks06} and stishovite\cite{Millot2015a}. 
A strong dependence in temperature is unveiled: the curves are almost indistinguishable. In particular the small difference between the quartz (2.65 g/cm$^3$) and the stishovite (4.29 g/cm$^3$) suggests that changes in the temperature dependence of the reflectivity onset induced by the slight density increase for precompressed quartz is most likely negligible given the experimental uncertainties. This is in good agreement with recent DFT-MD results suggesting that the shock reflectivity for precompressed quartz up to 2.94 g/cm$^3$ (P$_0$=5 GPa) depends only on the shock temperature\cite{Qi2015}. 

Using the experimentally determined R(U$_S$) and T(U$_S$) for quartz, we combine them in a reference curve:
\begin{equation}
R_{fit}(2.65, T)=\frac{0.11}{1+(16968/T)^{3.64}}T^{0.095}
\end{equation}
Then, using the previously described parametrization to compute the shock temperature of the precompressed quartz T($\rho_{0}$,U$_S$) we obtain the reflectivity of the shocked precompressed quartz:
\begin{equation}
	R(\rho_{0})=R_{fit}(2.65, T(\rho_{0}))
\end{equation}

\begin{figure}[htp]
	\centerline{\includegraphics[width=8cm,trim=0 0 0 0,clip]{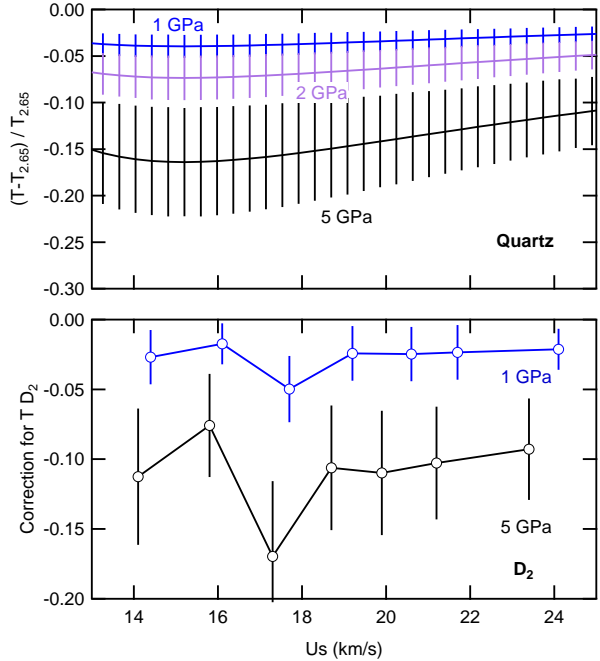}}
\caption{Magnitude of the correction for the temperature of quartz (T-T$_{2.65}$)/T$_{2.65}$ (upper figure) and for D$_{2}$ (lower figure) at P$_0$=1 and 5 GPa. The incertainties are the systematic uncertainties coming only from the uncertainties in $\beta$(U$_S$).} \label{fig:CorrectionTemp}
\end{figure}

\begin{figure}[htp]
	\centerline{\includegraphics[width=8cm,trim=0 0 0 0,clip]{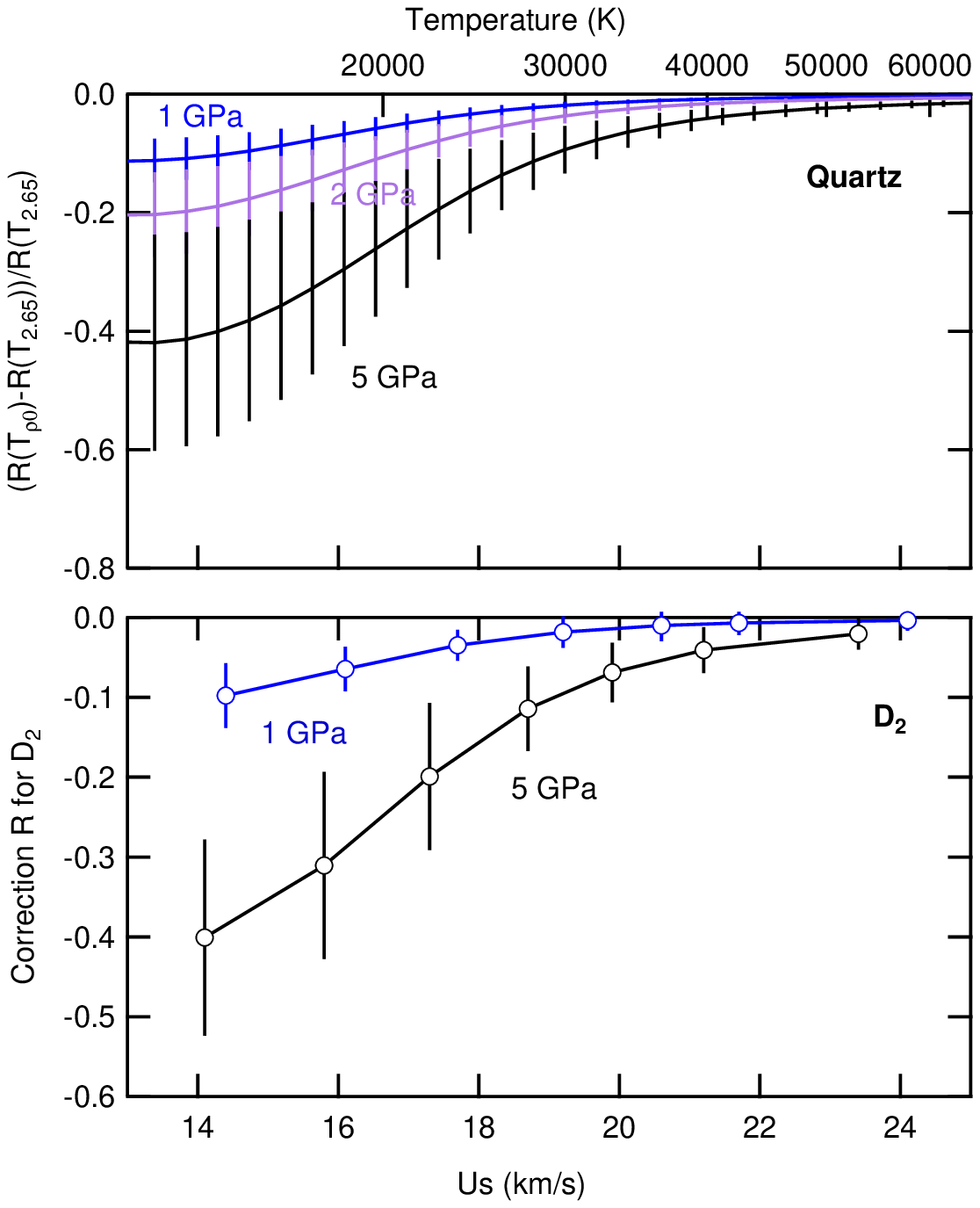}}
\caption{Magnitude of the correction for the reflectivity of quartz (R-R$_{2.65}$)/R$_{2.65}$ (upper figure) and for D2 (lower figure) at P$_0$=1 and 5 GPa. The incertainties are the systematic uncertainties coming only from the uncertainties in T$_{qz}$.} \label{fig:RvsUs_err}
\end{figure}
  
The impact on the corrections in temperature and reflectivity for D$_2$ defined as 
\begin{align}
\frac{T^{D2}_{\rho_{qz}=\rho_{0}}-T^{D2}_{\rho_{qz}=2.65}}{T^{D2}_{\rho_{qz}=2.65}}
\end{align}
and
\begin{align}
\frac{R^{D2}_{\rho_{qz}=\rho_{0}}-R^{D2}_{\rho_{qz}=2.65}}{R^{D2}_{\rho_{qz}=2.65}}
\end{align}
 obtained using this model are presented in Figures~\ref{fig:CorrectionTemp} and \ref{fig:RvsUs_err}.
An initial density of 2.94 g/cm$^3$ (5 GPa) gives a temperature correction for the quartz  $\sim$ 15-20\% which gives for the reflectivity a correction that can be as high as 60\% at low shock velocities. 

Consequently, the reflectivity corrections for D$_2$ (lower Figure~\ref{fig:RvsUs_err}) are of the same order of magnitude. A 15\% temperature correction (lower Figure~\ref{fig:CorrectionTemp}) is observed for a 5 GPa precompression. Note that both the precompression correction on the quartz temperature and on the quartz reflectivity affect the final inferred sample temperature.

\subsubsection{Re-analysis of helium, deuterium and hydrogen shock reflectivity and temperature data}

\begin{figure}[htp]
\includegraphics[width=80mm,trim=0 0 0 0,clip]{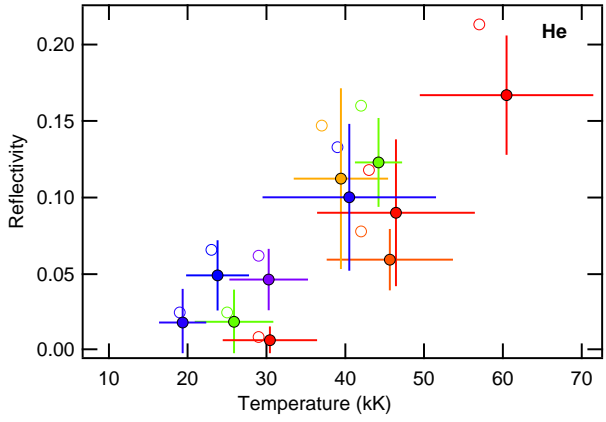}
\caption{Helium shock reflectivity as a function of the shock temperature: open and solid symbols are respectively the published data\cite{Celliers2010} and the re-analyzed ones. The colors show the initial pressure as in Figure ~\ref{fig:figHeb}.} \label{fig:figRvsTHe}
\end{figure}

\begin{figure}[htp]
\includegraphics[width=80mm,trim=0 0 0 0,clip]{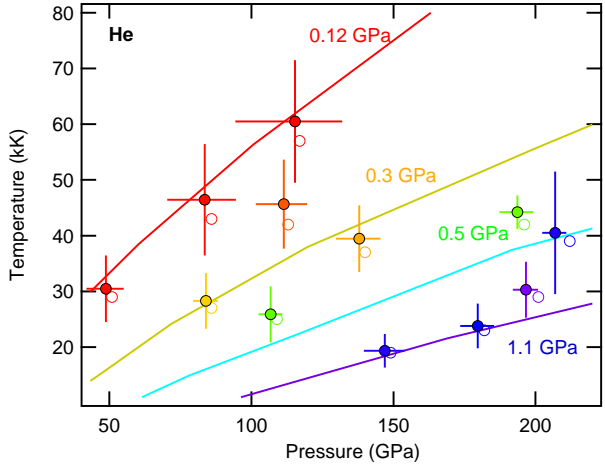}
\caption{Helium shock temperature as a function of the shock pressure: open and solid symbols are respectively the published data\cite{Celliers2010} and the re-analyzed ones. The initial pressure indicated in colors correspond to the experimental initial pressure. DFT-MD simulations\cite{Militzer} for 0.12 GPa (red), 0.35 GPa (orange), 0.85 GPa (light blue) and 1.8 GPa (purple)are also presented (solid lines).} \label{fig:figTvsPHe}
\end{figure}

\begin{figure}[htp]
\includegraphics[width=80mm,trim=0 0 0 0,clip]{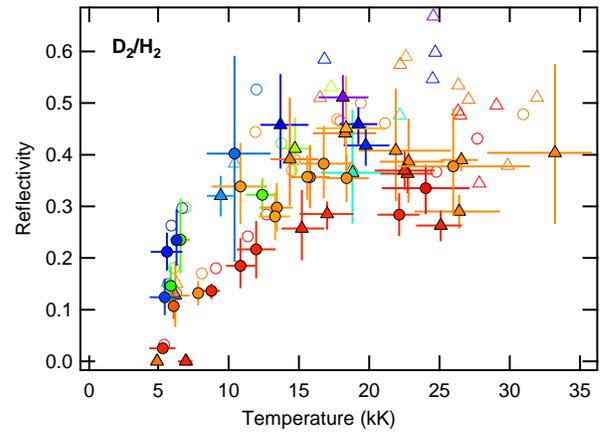}
\caption{Hydrogen and deuterium shock reflectivity as a function of the shock temperature: open and solid symbols are respectively the published data\cite{loubeyre} and the re-analyzed ones. Triangles are D$_{2}$ data and circles H$_{2}$ data. The colors show the initial pressure as in Figure ~\ref{fig:figD2H2}. } \label{fig:figRvsTH2}
\end{figure}

\begin{figure}[htp]
\includegraphics[width=80mm,trim=0 0 0 0,clip]{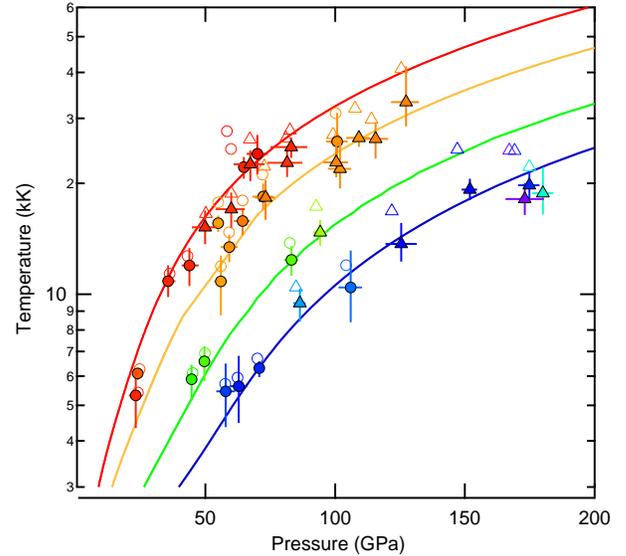}
\caption{Hydrogen and deuterium shock temperature as a function of the shock pressure: open and solid symbols are respectively the published data\cite{loubeyre} and the re-analyzed ones. Triangles are D$_{2}$ data and circles H$_{2}$ data. The solid lines are DFT-MD calculations from Caillabet et al.\cite{Caillabet}. The colors show the initial pressure as in Figure ~\ref{fig:figD2H2}.} \label{fig:figTvsPH2}
\end{figure}

Figures~\ref{fig:figRvsTHe},~\ref{fig:figTvsPHe},~\ref{fig:figRvsTH2} and~\ref{fig:figTvsPH2} present a comparison between the published data on helium and hydrogen and the data re-analyzed with the new model. 

In the previous reports\cite{Celliers2010,loubeyre}, the precompression was accounted for with a slightly different approach:  the temperature shift expected for a precompressed Hugoniot was estimated using a Gruneisen model\cite{TempGrun} calibrated against the difference in shock temperature at a given pressure between fused silica and quartz. Similarly, the influence on the reflectivity was modeled based on the observed difference in onset and maximum reflectivity between fused silica and quartz. 

At the relatively modest precompression achieved in these previously published datasets, the new precompression correction does not strongly affect the results. Instead, the changes observed for the temperature and the reflectivity are mainly arising from the fit used for the quartz reflectivity experimental data along the quartz principal Hugoniot\cite{Celliers2010}. 

The change in helium shock temperature are small but the reflectivity appears lower than previously thought. This suggest lower electronic conductivities in the explored temperature-density domain. We observe that reflectivity saturation has not been reached yet but on-going measurements aiming at higher pressures and densities might reach the expected reflectivity saturation. 

For Hydrogen, as for helium, the decrease in reflectivity reduces the temperature and yields a better agreement between the data and the DFT-MD calculations\cite{Caillabet} above 10$^4$ K.

\section{Conclusion}
 Laser driven shocks on precompressed samples allow reaching completely uncharted territories in the phase diagram of low-Z system and provide solid benchmarks for advanced warm dense matter theories and planetary science models, but require to carefully account for the precompression of the quartz reference, in particular when the precompression exceeds a few GPa. The essential calibration fits and analytical corrections to use quartz as a standard are summarized in Table~\ref{tab:summary}. Future refinements of the quartz standard, and in particular knowledge of the release behavior, should improve the precision and accuracy of the past and future experiments. 

\begin{table*}
 \centering     
 \caption{Summary of the equations and fits needed to use quartz as a standard for pressure, density, reflectivity and temperature relative measurements. Velocities are in km/s and temperature in K.}
     \begin{tabular}{l l}
\hline
\hline &\\
\vspace{0.2cm} Initial density: $\rho_{0}$ & $\rho_{0}$=2.649(P$_{0}$(4.9/37.7)+1)$^{1/4.9}$\\
\vspace{0.2cm}Index of refraction: n$_{Q0}$ & n$_{Q0}$=1.54687+0.1461(1)($\rho_0$-2.649)\\
      \hline &\\
 Hugoniot: U$_{S}$(2.65,u$_{p}$)&U$_S$(u$_{p}<$ d )=a + b u$_{p}$ - c u$_{p}^{2}$ \\
 
Piecewise polynomial fit & U$_S$(u$_{p}\geq$ d )=(a + c d$^{2}$) +(b -2 c d) u$_{p}$ \\
 & a=2.124~$\pm$~0.121 ;  b=1.7198~$\pm$~0.0258\\
\vspace{0.3cm}  &  c=0.01744~$\pm$~0.00133 ; d=14.168~$\pm$~0.731 \\
Hugoniot shift with initial density : &  U$_{S}$=U$_{S}$(2.65,u$_{p}$)+$\alpha$ ($\rho_{0}$-2.65)  \\
  \vspace{0.2cm}  &  $\alpha$= 2.3($\pm$0.4) - 0.037 ($\pm$0.027)u$_p$\\
\hline &\\
\vspace{0.3cm}Temperature quartz: T$(2.65,U_S) $ & T$_{qz}$(K) =1860($\pm$ 190) + 3.56($\pm$ 0.52)U$_{S}$(km/s)$^{3.036(\pm)0.046}$  \\
  \vspace{0.2cm}Temperature shift:  & T($\rho_{0},U_S)$=T(2.65,Us)-$\beta$ ($\rho_{0}$-2.65) \\
  \vspace{0.2cm} & $\beta\pm30\%=-14786 + 1555 ~ U_S$\\
\hline &\\
\vspace{0.3cm}Reflectivity quartz: R(2.65,T)& R(2.65,T) = $\frac{\displaystyle 0.11(\pm0.03)}{\displaystyle 1+(16968(\pm737)/T)^{3.64(\pm0.51)}}$T$^{0.095(\pm0.029)}$\\
\vspace{0.2cm}Reflectivity shift: & R($\rho_{0}$) = R(2.65,T($\rho_{0}$,Us))\\
\hline\hline

   \end{tabular}

     \label{tab:summary}

 \end{table*}

\begin{table*}
 \centering
 
 \caption{He Hugoniot data: pressure (P), compression ratio ($\rho / \rho_{0}$), temperature, reflectivity and variation of internal energy E-E$_{0}$ using the present model for the quartz reference and the shock velocity data from Eggert et al\cite{Eggert08} and Celliers et al\cite{Celliers2010}. For all quantities, the total errors (which include the systematic errors due to the quartz standard and random errors due to experiment) are given in parentheses and correspond to the error on the last digits. For example: $\rho/\rho_{0}$(-,+) = 4.97 (58,76) means $\rho/\rho_{0}min$=4.97-0.58 and $\rho/\rho_{0}max$=4.97+0.76  }
     \begin{tabular}{cccccc}

   \hline  \hline 
      Shot	& \hspace {0.1cm}	P (GPa)	(-,+) & \hspace {0.1cm}	$\rho/\rho_{0}$(-,+)	& \hspace {0.1cm}	T (kK) (-,+)	& \hspace {0.1cm}	R	(-,+)& \hspace {0.1cm}	E-E$_{0}$ (kJ/g) (-,+)	\\
 \hline  
33488	& \hspace {0.1cm}	99 (3,3)	& \hspace {0.1cm}	1.89 (6,6)	& \hspace {0.1cm}	8.8 (0.4,0.3)	& \hspace {0.1cm}	0.0 (1,1)	& \hspace {0.1cm}	57 (3,3)	\\
34833	& \hspace {0.1cm}	111 (10,8)	& \hspace {0.1cm}	4.96 (62,82)	& \hspace {0.1cm}	45.7 (3.2,2.7)	& \hspace {0.1cm}	0.6 (2,2)	& \hspace {0.1cm}	251 (11,10)	\\
34836	& \hspace {0.1cm}	107 (4,4)	& \hspace {0.1cm}	3.06 (24,27)	& \hspace {0.1cm}	25.9 (1.6,1.4)	& \hspace {0.1cm}	0.2 (2,2)	& \hspace {0.1cm}	122 (5,5)	\\
36177	& \hspace {0.1cm}	84 (5,4)	& \hspace {0.1cm}	3.41 (26,29)	& \hspace {0.1cm}	28.3 (1.7,1.4)	& \hspace {0.1cm}	0.0 (1,1)	& \hspace {0.1cm}	131 (5,5)	\\
36178	& \hspace {0.1cm}	138 (8,7)	& \hspace {0.1cm}	4.34 (43,53)	& \hspace {0.1cm}	39.5 (2.4,1.8)	& \hspace {0.1cm}	0.11 (6,6)	& \hspace {0.1cm}	247 (10,9)	\\
38993	& \hspace {0.1cm}	115 (21,17)	& \hspace {0.1cm}	5.59 (64,92)	& \hspace {0.1cm}	60.5 (3.3,2.7)	& \hspace {0.1cm}	0.17 (4,4)	& \hspace {0.1cm}	388 (17,16)	\\
38994	& \hspace {0.1cm}	84 (13,11)	& \hspace {0.1cm}	5.48 (70,99)	& \hspace {0.1cm}	46.4 (3.3,2.8)	& \hspace {0.1cm}	0.09 (5,5)	& \hspace {0.1cm}	263 (11,11)	\\
40131	& \hspace {0.1cm}	197 (5,4)	& \hspace {0.1cm}	3.40 (20,19)	& \hspace {0.1cm}	30.3 (3.0,2.9)	& \hspace {0.1cm}	0.05 (2,2)	& \hspace {0.1cm}	163 (6,6)	\\
41452	& \hspace {0.1cm}	147 (7,7)	& \hspace {0.1cm}	2.94 (32,43)	& \hspace {0.1cm}	19.4 (0.9,0.7)	& \hspace {0.1cm}	0.02 (2,2)	& \hspace {0.1cm}	117 (5,5)	\\
41453	& \hspace {0.1cm}	194 (6,6)	& \hspace {0.1cm}	3.88 (24,24)	& \hspace {0.1cm}	44.2 (1.8,1.3)	& \hspace {0.1cm}	0.12 (3,3)	& \hspace {0.1cm}	246 (9,9)	\\
41454	& \hspace {0.1cm}	180 (6,6)	& \hspace {0.1cm}	2.90 (22,23)	& \hspace {0.1cm}	23.8 (1.2,0.8)	& \hspace {0.1cm}	0.05 (2,2)	& \hspace {0.1cm}	144 (6,6)	\\
41455	& \hspace {0.1cm}	49 (8,6)	& \hspace {0.1cm}	4.52 (72,91)	& \hspace {0.1cm}	30.4 (2.6,2.6)	& \hspace {0.1cm}	0.01 (1,1)	& \hspace {0.1cm}	146 (7,7)	\\
43301	& \hspace {0.1cm}	207 (4,4)	& \hspace {0.1cm}	3.34 (17,15)	& \hspace {0.1cm}	40.5 (5.3,5.9)	& \hspace {0.1cm}	0.10 (5,5)	& \hspace {0.1cm}	176 (6,6)	\\
\hline \hline 

   \end{tabular}
     \label{tab:He}

 \end{table*}
 
\begin{table*}
 \centering
      \caption{H$_{2}$ Hugoniot data: pressure (P), compression ratio ($\rho / \rho_{0}$), temperature, reflectivity and variation of internal energy E-E$_{0}$ using the present model for the quartz reference and the shock velocity data given in Table I of Loubeyre et al\cite{loubeyre}. For all quantities, the total errors (which include the systematic errors due to the quartz standard and random errors due to experiment) are given in parentheses and correspond to the error on the last digits. For example: $\rho/\rho_{0}$(-,+) = 4.97 (58,76) means $\rho/\rho_{0}min$=4.97-0.58 and $\rho/\rho_{0}max$=4.97+0.76 For the shots in the lower part of the table, the shock velocity in H$_{2}$ could only be obtained with transit time measurement hence not accurately enough for determining the compression ratio or the energy variation}
     \begin{tabular}{cccccc}

    \hline \hline 
      Shot	& \hspace {0.1cm}	P (GPa)	(-,+) & \hspace {0.1cm}	$\rho/\rho_{0}$(-,+)	& \hspace {0.1cm}	T (kK) (-,+)	& \hspace {0.1cm}	R	(-,+)& \hspace {0.1cm}	E-E$_{0}$ (kJ/g) (-,+)	\\
 \hline 
34834	& \hspace {0.1cm}	56 (3,2)	& \hspace {0.1cm}	4.40 (37,40)	& \hspace {0.1cm}	10.8 (2.0,1.9)	& \hspace {0.1cm}	0.34 (7,9)	& \hspace {0.1cm}	247 (11,10)	\\
34835	& \hspace {0.1cm}	24 (2,2)	& \hspace {0.1cm}	3.85 (36,39)	& \hspace {0.1cm}	6.1 (0.2,0.2)	& \hspace {0.1cm}	0.11 (3,3)	& \hspace {0.1cm}	118 (6,6)	\\
36174	& \hspace {0.1cm}	50 (2,2)	& \hspace {0.1cm}	3.87 (35,41)	& \hspace {0.1cm}	6.6 (0.8,0.7)	& \hspace {0.1cm}	0.24 (6,8)	& \hspace {0.1cm}	154 (8,8)	\\
36176	& \hspace {0.1cm}	45 (2,2)	& \hspace {0.1cm}	3.64 (34,40)	& \hspace {0.1cm}	5.9 (0.7,0.6)	& \hspace {0.1cm}	0.15 (4,4)	& \hspace {0.1cm}	133 (9,9)	\\
38326	& \hspace {0.1cm}	70 (4,4)	& \hspace {0.1cm}	4.91 (54,64)	& \hspace {0.1cm}	24.0 (3.1,3.0)	& \hspace {0.1cm}	0.34 (5,5)	& \hspace {0.1cm}	486 (54,63)	\\
38991	& \hspace {0.1cm}	83 (3,3)	& \hspace {0.1cm}	4.11 (35,39)	& \hspace {0.1cm}	12.4 (1.2,1.1)	& \hspace {0.1cm}	0.32 (4,4)	& \hspace {0.1cm}	264 (14,14)	\\
38997	& \hspace {0.1cm}	44 (3,4)	& \hspace {0.1cm}	5.09 (70,79)	& \hspace {0.1cm}	12.0 (1.3,1.4)	& \hspace {0.1cm}	0.22 (5,6)	& \hspace {0.1cm}	275 (26,36)	\\
39000	& \hspace {0.1cm}	63 (3,3)	& \hspace {0.1cm}	3.55 (44,55)	& \hspace {0.1cm}	5.6 (1.2,1.2)	& \hspace {0.1cm}	0.21 (4,4)	& \hspace {0.1cm}	147 (14,15)	\\
41451	& \hspace {0.1cm}	58 (4,4)	& \hspace {0.1cm}	3.10 (37,48)	& \hspace {0.1cm}	5.5 (1.0,1.1)	& \hspace {0.1cm}	0.12 (3,4)	& \hspace {0.1cm}	129 (15,15)	\\
41458	& \hspace {0.1cm}	23 (3,2)	& \hspace {0.1cm}	4.30 (64,88)	& \hspace {0.1cm}	5.3 (0.10,0.9)	& \hspace {0.1cm}	0.03 (1,1)	& \hspace {0.1cm}	139 (8,8)	\\
43297	& \hspace {0.1cm}	101 (4,4)	& \hspace {0.1cm}	4.38 (26,27)	& \hspace {0.1cm}	26.0 (4.0,5.2)	& \hspace {0.1cm}	0.38 (9,10)	& \hspace {0.1cm}	449 (15,14)	\\
43298	& \hspace {0.1cm}	106 (5,4)	& \hspace {0.1cm}	3.58 (39,47)	& \hspace {0.1cm}	10.4 (1.10,2.6)	& \hspace {0.1cm}	0.40 (21,19)	& \hspace {0.1cm}	252 (20,20)	\\
47716	& \hspace {0.1cm}	36 (2,3)	& \hspace {0.1cm}	4.93 (48,47)	& \hspace {0.1cm}	10.8 (1.0,1.1)	& \hspace {0.1cm}	0.18 (4,5)	& \hspace {0.1cm}	222 (11,15)	\\
47719	& \hspace {0.1cm}	64 (3,3)	& \hspace {0.1cm}	4.57 (33,37)	& \hspace {0.1cm}	15.8 (1.3,1.3)	& \hspace {0.1cm}	0.36 (6,6)	& \hspace {0.1cm}	305 (12,11)	\\
52250	& \hspace {0.1cm}	55 (2,2)	& \hspace {0.1cm}	4.44 (30,33)	& \hspace {0.1cm}	15.6 (0.9,0.8)	& \hspace {0.1cm}	0.36 (4,4)	& \hspace {0.1cm}	240 (8,8)	\\
53835	& \hspace {0.1cm}	59 (3,3)	& \hspace {0.1cm}	4.27 (57,68)	& \hspace {0.1cm}	13.4 (1.1,1.1)	& \hspace {0.1cm}	0.30 (5,5)	& \hspace {0.1cm}	263 (19,19)	\\
53838	& \hspace {0.1cm}	65 (2,2)	& \hspace {0.1cm}	4.97 (31,29)	& \hspace {0.1cm}	22.1 (1.6,1.4)	& \hspace {0.1cm}	0.28 (4,4)	& \hspace {0.1cm}	404 (29,33)	\\
55003	& \hspace {0.1cm}	72 (4,4)	& \hspace {0.1cm}	4.32 (52,67)	& \hspace {0.1cm}	18.4 (1.5,1.6)	& \hspace {0.1cm}	0.35 (4,5)	& \hspace {0.1cm}	321 (22,21)	\\
56366	& \hspace {0.1cm}	71 (2,2)	& \hspace {0.1cm}	3.37 (19,19)	& \hspace {0.1cm}	6.3 (0.3,0.3)	& \hspace {0.1cm}	0.23 (5,6)	& \hspace {0.1cm}	162 (7,6)	\\
			& \hspace {0.1cm}		& \hspace {0.1cm}		& \hspace {0.1cm}		& \hspace {0.1cm}		\\
50377	& \hspace {0.1cm}		& \hspace {0.1cm}		& \hspace {0.1cm}	16.8 (1.8,1.8)	& \hspace {0.1cm}	0.38 (6,7)	& \hspace {0.1cm}		\\
53471	& \hspace {0.1cm}		& \hspace {0.1cm}		& \hspace {0.1cm}	8.8 (0.7,0.6)	& \hspace {0.1cm}	0.14 (2,1)	& \hspace {0.1cm}		\\
53472	& \hspace {0.1cm}		& \hspace {0.1cm}		& \hspace {0.1cm}	7.8 (0.4,0.4)	& \hspace {0.1cm}	0.13 (2,2)	& \hspace {0.1cm}		\\
53478	& \hspace {0.1cm}		& \hspace {0.1cm}		& \hspace {0.1cm}	13.3 (1.2,1.1)	& \hspace {0.1cm}	0.28 (5,4)	& \hspace {0.1cm}		\\

\hline \hline 

   \end{tabular}

     \label{tab:H2}

 \end{table*}

\begin{table*}
 \centering
 
 \caption{D$_{2}$ Hugoniot data: pressure (P), compression ratio ($\rho / \rho_{0}$), temperature, reflectivity and variation of internal energy E-E$_{0}$ using the present model for the quartz reference and the shock velocity data given in Table I of Loubeyre et al\cite{loubeyre}. For all quantities, the total errors (which include the systematic errors due to the quartz standard and random errors due to experiment) are given in parentheses and correspond to the error on the last digits. For example: $\rho/\rho_{0}$(-,+) = 4.97 (58,76) means $\rho/\rho_{0}min$=4.97-0.58 and $\rho/\rho_{0}max$=4.97+0.76 For the shots in the lower part of the table, the shock velocity in D$_{2}$ could only be obtained with transit time measurement hence not accurately enough for determining the compression ratio or the energy variation}
     \begin{tabular}{cccccc}

      \hline \hline 
      Shot	& \hspace {0.1cm}	P (GPa)	(-,+) & \hspace {0.1cm}	$\rho/\rho_{0}$(-,+)	& \hspace {0.1cm}	T (kK) (-,+)	& \hspace {0.1cm}	R	(-,+)& \hspace {0.1cm}	E-E$_{0}$ (kJ/g) (-,+)	\\
 \hline    
40133	& \hspace {0.1cm}	116 (6,5)	& \hspace {0.1cm}	3.95 (25,27)	& \hspace {0.1cm}	26.4 (3.2,3.0)	& \hspace {0.1cm}	0.29 (3,3)	& \hspace {0.1cm}	254 (10,10)	\\
41449	& \hspace {0.1cm}	102 (5,4)	& \hspace {0.1cm}	4.45 (35,39)	& \hspace {0.1cm}	21.9 (2.5,3.8)	& \hspace {0.1cm}	0.41 (10,13)	& \hspace {0.1cm}	225 (9,9)	\\
41459	& \hspace {0.1cm}	73 (4,4)	& \hspace {0.1cm}	4.29 (36,40)	& \hspace {0.1cm}	18.3 (2.4,2.2)	& \hspace {0.1cm}	0.44 (2,1)	& \hspace {0.1cm}	168 (8,8)	\\
47715	& \hspace {0.1cm}	105 (3,3)	& \hspace {0.1cm}	3.79 (30,31)	& \hspace {0.1cm}	2.3 (0.1,0.1)	& \hspace {0.1cm}	0.36 (10,12)	& \hspace {0.1cm}	131 (7,6)	\\
47718	& \hspace {0.1cm}	180 (4,4)	& \hspace {0.1cm}	4.03 (25,27)	& \hspace {0.1cm}	18.8 (2.3,3.3)	& \hspace {0.1cm}	0.36 (10,13)	& \hspace {0.1cm}	236 (9,9)	\\
47720	& \hspace {0.1cm}	100 (5,5)	& \hspace {0.1cm}	4.17 (30,35)	& \hspace {0.1cm}	22.8 (2.3,2.5)	& \hspace {0.1cm}	0.39 (8,8)	& \hspace {0.1cm}	225 (10,9)	\\
47721	& \hspace {0.1cm}	50 (5,4)	& \hspace {0.1cm}	4.95 (52,69)	& \hspace {0.1cm}	15.2 (1.5,1.6)	& \hspace {0.1cm}	0.26 (6,7)	& \hspace {0.1cm}	154 (8,8)	\\
50369	& \hspace {0.1cm}	94 (3,2)	& \hspace {0.1cm}	3.72 (22,25)	& \hspace {0.1cm}	14.7 (1.1,1.2)	& \hspace {0.1cm}	0.41 (5,6)	& \hspace {0.1cm}	151 (6,6)	\\
50370	& \hspace {0.1cm}	86 (3,2)	& \hspace {0.1cm}	3.49 (24,27)	& \hspace {0.1cm}	9.5 (1.0,0.9)	& \hspace {0.1cm}	0.32 (4,4)	& \hspace {0.1cm}	102 (5,5)	\\
50372	& \hspace {0.1cm}	82 (8,6)	& \hspace {0.1cm}	4.90 (44,53)	& \hspace {0.1cm}	22.7 (1.8,1.8)	& \hspace {0.1cm}	0.36 (4,3)	& \hspace {0.1cm}	242 (10,10)	\\
50378	& \hspace {0.1cm}	125 (6,6)	& \hspace {0.1cm}	3.71 (45,58)	& \hspace {0.1cm}	13.7 (1.4,1.9)	& \hspace {0.1cm}	0.46 (8,10)	& \hspace {0.1cm}	144 (12,12)	\\
52253	& \hspace {0.1cm}	152 (3,3)	& \hspace {0.1cm}	3.57 (19,18)	& \hspace {0.1cm}	19.2 (1.3,1.2)	& \hspace {0.1cm}	0.46 (4,3)	& \hspace {0.1cm}	172 (6,5)	\\
53473	& \hspace {0.1cm}	109 (5,4)	& \hspace {0.1cm}	4.10 (25,27)	& \hspace {0.1cm}	26.6 (1.5,1.2)	& \hspace {0.1cm}	0.39 (2,2)	& \hspace {0.1cm}	235 (8,7)	\\
53474	& \hspace {0.1cm}	173 (7,7)	& \hspace {0.1cm}	3.61 (38,51)	& \hspace {0.1cm}	18.1 (1.8,1.8)	& \hspace {0.1cm}	0.51 (5,5)	& \hspace {0.1cm}	189 (14,14)	\\
53839	& \hspace {0.1cm}	127 (6,5)	& \hspace {0.1cm}	4.12 (26,27)	& \hspace {0.1cm}	33.2 (4.10,8.5)	& \hspace {0.1cm}	0.40 (13,17)	& \hspace {0.1cm}	279 (10,9)	\\
56360	& \hspace {0.1cm}	175 (4,4)	& \hspace {0.1cm}	3.71 (24,24)	& \hspace {0.1cm}	19.8 (1.6,1.7)	& \hspace {0.1cm}	0.42 (4,4)	& \hspace {0.1cm}	203 (9,8)	\\
56370	& \hspace {0.1cm}	83 (8,6)	& \hspace {0.1cm}	4.79 (41,47)	& \hspace {0.1cm}	25.1 (1.6,1.6)	& \hspace {0.1cm}	0.26 (3,3)	& \hspace {0.1cm}	253 (9,9)	\\
55005	& \hspace {0.1cm}	60 (6,5)	& \hspace {0.1cm}	4.70 (73,96)	& \hspace {0.1cm}	17.0 (1.9,1.9)	& \hspace {0.1cm}	0.29 (3,3)	& \hspace {0.1cm}	177 (16,16)	\\
58084	& \hspace {0.1cm}	67 (6,6)	& \hspace {0.1cm}	4.74 (32,37)	& \hspace {0.1cm}	22.5 (2.3,2.3)	& \hspace {0.1cm}	0.37 (4,3)	& \hspace {0.1cm}	205 (7,6)	\\
			& \hspace {0.1cm}		& \hspace {0.1cm}		& \hspace {0.1cm}		& \hspace {0.1cm}		\\
47723	& \hspace {0.1cm}		& \hspace {0.1cm}		& \hspace {0.1cm}	18.4 (2.1,2.7)	& \hspace {0.1cm}	0.45 (8,10)	& \hspace {0.1cm}		\\
50371	& \hspace {0.1cm}		& \hspace {0.1cm}		& \hspace {0.1cm}	6.9 (0.6,0.5)	& \hspace {0.1cm}	0.0 (0,0)	& \hspace {0.1cm}		\\
50374	& \hspace {0.1cm}		& \hspace {0.1cm}		& \hspace {0.1cm}	4.9 (0.4,0.3)	& \hspace {0.1cm}	0.0 (0,0)	& \hspace {0.1cm}		\\
50376	& \hspace {0.1cm}		& \hspace {0.1cm}		& \hspace {0.1cm}	6.2 (0.9,0.10)	& \hspace {0.1cm}	0.13 (6,7)	& \hspace {0.1cm}		\\
50381	& \hspace {0.1cm}		& \hspace {0.1cm}		& \hspace {0.1cm}	14.4 (1.4,2.2)	& \hspace {0.1cm}	0.39 (9,13)	& \hspace {0.1cm}		\\

\hline \hline 

   \end{tabular}
     \label{tab:D2}

 \end{table*} 
 
\section*{Acknowledgments}
Authors gratefully acknowledge discussions with Stewart Mc Williams, Florent Occelli and Dylan Spaulding. Part of this work was prepared by LLNL under Contract DE-AC52-07NA27344 with support from LLNL LDRD program and the US Department of Energy through the joint FES/NNSA HEDLP program.

\appendix
\section{Characterization of the initial pre-compressed state }\label{Appendix:Static}

\subsection{Initial density}
The initial mass density is an important parameter for shock compression experiments. Here we use the experimentally determined equations of state at room temperature in the GPa pressure range to infer the initial density of the sample and the quartz reference plate.

The high pressure static compression of silica polymorphs has been extensively studied. $\alpha-quartz$, the stable phase at ambient conditions can be compressed to almost 20 GPa at room temperature without any phase transition, despite the fact that high pressure polymorphs coesite and stishovite become the stable phases above respectively 2 and 8 GPa \cite{Gregoryanz2000a}. The density of quartz at ambient conditions is $\rho_{00}=2.649\:g/cm^3$ \cite{Bottom1982,Barron1982}. Accurate ultrasound measurements up to 1 GPa\cite{Calderon2007} determined $\alpha-quartz$ room temperature bulk modulus $B_0$=37.5(0.2) GPa and its first derivative $B_0'$=4.7(0.5), data that can be used directly in the Birch-Murnaghan equation that determines the final density as a function of pressure. These values are in very good agreement with x-ray diffraction experiments and state-of-the-art \emph{ab-initio} simulations to pressure above 20 GPa giving $B_0$=37.7(3) GPa and its first derivative $B_0'$=4.9(1)\cite{Kimizuka2007}. The relative uncertainty on the initial density $\Delta \rho_0/\rho_0$ is then dominated by the uncertainty on $B_0'$ and is 0.3$\%$ at 6 GPa and 0.6$\%$ at 10 GPa.

\subsection{Refractive index}
VISAR measurements of reflecting shock velocity in transparent media actually measure an apparent velocity $U_{S~app}=n_0 U_S$. In order to obtain the shock velocity U$_S$ we therefore need to know $n_0$ the refractive index at the VISAR wavelength (532 nm on Omega), room temperature and initial pressure $P_0$ for both the reference and the sample.  

We use z-cut $\alpha$-quartz plates. At ambient conditions, the refractive index for the ordinary rays of quartz crystal \cite{Ghosh1999} at 532 nm is $n_{Q00,532}=1.54687$. The refractive index of quartz is known to increase with pressure. We updated the index pressure variation using literature experimental data\cite{Vedam1967} and recent elastic constants\cite{Calderon2007,noteIndex}. A linear fit as a function of density gives the index of precompressed quartz $n_{Q0}=n_{Q00}+0.1461(1)(\rho_0-\rho_{00})$.


\begin{thebibliography}{45}
\expandafter\ifx\csname
natexlab\endcsname\relax\def\natexlab#1{#1}\fi
\expandafter\ifx\csname bibnamefont\endcsname\relax
  \def\bibnamefont#1{#1}\fi
\expandafter\ifx\csname bibfnamefont\endcsname\relax
  \def\bibfnamefont#1{#1}\fi
\expandafter\ifx\csname citenamefont\endcsname\relax
  \def\citenamefont#1{#1}\fi
\expandafter\ifx\csname url\endcsname\relax
  \def\url#1{\texttt{#1}}\fi
\expandafter\ifx\csname
urlprefix\endcsname\relax\def\urlprefix{URL }\fi
\providecommand{\bibinfo}[2]{#2}
\providecommand{\eprint}[2][]{\url{#2}}


\bibitem[{\citenamefont{Loubeyre}(2003)}]{Loubeyre03}
\bibinfo{author}{\bibfnamefont{P.} \bibnamefont{Loubeyre}},
\bibinfo{author}{\bibfnamefont{P. M.} \bibnamefont{Celliers}},
\bibinfo{author}{\bibfnamefont{D. G.} \bibnamefont{Hicks}},
\bibinfo{author}{\bibfnamefont{E.} \bibnamefont{Henry}},
\bibinfo{author}{\bibfnamefont{A.} \bibnamefont{Dewaele}},
\bibinfo{author}{\bibfnamefont{J.} \bibnamefont{Pasley}},
\bibinfo{author}{\bibfnamefont{J.} \bibnamefont{Eggert}},
\bibinfo{author}{\bibfnamefont{M.} \bibnamefont{Koenig}},
\bibinfo{author}{\bibfnamefont{F.} \bibnamefont{Occeli}},
\bibinfo{author}{\bibfnamefont{K. M.} \bibnamefont{Lee}},
\bibinfo{author}{\bibfnamefont{R.} \bibnamefont{Jeanloz}},
\bibinfo{author}{\bibfnamefont{D.} \bibnamefont{Neely}},
\bibinfo{author}{\bibfnamefont{A.} \bibnamefont{Benuzzi-Mounaix}},
\bibinfo{author}{\bibfnamefont{D.} \bibnamefont{Bradley}},
\bibinfo{author}{\bibfnamefont{M.} \bibnamefont{Bastea}},
\bibinfo{author}{\bibfnamefont{S.} \bibnamefont{Moon}},
\bibinfo{author}{\bibfnamefont{G.} \bibnamefont{Collins}},
  \bibinfo{journal}{High Pres. Res.} \textbf{\bibinfo{volume}{24}},
  \bibinfo{pages}{25} (\bibinfo{year}{2004}).

\bibitem[{\citenamefont{Jeanloz}(2007)}]{Jeanloz07}
\bibinfo{author}{\bibfnamefont{R.} \bibnamefont{Jeanloz}},
\bibinfo{author}{\bibfnamefont{P. M.} \bibnamefont{Celliers}},
\bibinfo{author}{\bibfnamefont{G. W.} \bibnamefont{Collins}},
\bibinfo{author}{\bibfnamefont{J. H.} \bibnamefont{Eggert}},
\bibinfo{author}{\bibfnamefont{K. K. M.} \bibnamefont{Lee}},
\bibinfo{author}{\bibfnamefont{R. S.} \bibnamefont{McWilliams}},
\bibinfo{author}{\bibfnamefont{S.} \bibnamefont{Brygoo}},
\bibinfo{author}{\bibfnamefont{P.} \bibnamefont{Loubeyre}},
  \bibinfo{journal}{PNAS} \textbf{\bibinfo{volume}{104}},
  \bibinfo{pages}{9172} (\bibinfo{year}{2007}).

\bibitem[{\citenamefont{Eggert}(2008)}]{Eggert08}
\bibinfo{author}{\bibfnamefont{J.} \bibnamefont{Eggert}},
\bibinfo{author}{\bibfnamefont{S.} \bibnamefont{Brygoo}},
\bibinfo{author}{\bibfnamefont{P.} \bibnamefont{Loubeyre}},
\bibinfo{author}{\bibfnamefont{R. S.} \bibnamefont{McWilliams}},
\bibinfo{author}{\bibfnamefont{P. M.} \bibnamefont{Celliers}},
\bibinfo{author}{\bibfnamefont{D. G.} \bibnamefont{Hicks}},
\bibinfo{author}{\bibfnamefont{T. R.} \bibnamefont{Boehly}},
\bibinfo{author}{\bibfnamefont{R.} \bibnamefont{Jeanloz}},
\bibinfo{author}{\bibfnamefont{G. W.} \bibnamefont{Collins}},
  \bibinfo{journal}{Phys. Rev. Let.} \textbf{\bibinfo{volume}{100}},
  \bibinfo{pages}{124503} (\bibinfo{year}{2008}).

\bibitem[{\citenamefont{Celliers}(2010)}]{Celliers2010}
\bibinfo{author}{\bibfnamefont{P. M.}\bibnamefont{Celliers}},
\bibinfo{author}{\bibfnamefont{P.}\bibnamefont{Loubeyre}}, 
\bibinfo{author}{\bibfnamefont{J. H.}\bibfnamefont{Eggert}},
\bibinfo{author}{\bibfnamefont{S.}\bibfnamefont{Brygoo}}, 
\bibinfo{author}{\bibfnamefont{R. S.}\bibfnamefont{McWilliams}},
\bibinfo{author}{\bibfnamefont{D. G.}\bibfnamefont{Hicks}},
\bibinfo{author}{\bibfnamefont{T. R.}\bibfnamefont{Boehly}},
\bibinfo{author}{\bibfnamefont{R.}\bibfnamefont{Jeanloz}},
\bibinfo{author}{\bibfnamefont{G. W.}\bibfnamefont{Collins}},
\bibinfo{journal}{Phys. Rev. Let.} \textbf{\bibinfo{volume}{104}},
\bibinfo{pages}{184503} (\bibinfo{year}{2010}).


\bibitem[{\citenamefont{Loubeyre}(2012)}]{loubeyre}
\bibinfo{author}{\bibfnamefont{P.} \bibnamefont{Loubeyre}},
\bibinfo{author}{\bibfnamefont{S.} \bibnamefont{Brygoo}},
\bibinfo{author}{\bibfnamefont{J.} \bibnamefont{Eggert}},
\bibinfo{author}{\bibfnamefont{P. M.} \bibnamefont{Celliers}},
\bibinfo{author}{\bibfnamefont{D. K.} \bibnamefont{Spaulding}},
\bibinfo{author}{\bibfnamefont{J. R.} \bibnamefont{Rygg}},
\bibinfo{author}{\bibfnamefont{T. R.} \bibnamefont{Boehly}},
\bibinfo{author}{\bibfnamefont{G. W.} \bibnamefont{Collins}},
\bibinfo{author}{\bibfnamefont{R.} \bibnamefont{Jeanloz}},
  \bibinfo{journal}{Phys. Rev. B.} \textbf{\bibinfo{volume}{86}},
  \bibinfo{pages}{144115} (\bibinfo{year}{2012}).  
  
\bibitem[{\citenamefont{Celliers}(2004)}]{Celliers04}
\bibinfo{author}{\bibfnamefont{P. M.} \bibnamefont{Celliers}},
\bibinfo{author}{\bibfnamefont{D. K.} \bibnamefont{Bradley}},
\bibinfo{author}{\bibfnamefont{G. W.} \bibnamefont{Collins}},
\bibinfo{author}{\bibfnamefont{D. G.} \bibnamefont{Hicks}},
\bibinfo{author}{\bibfnamefont{T. R.} \bibnamefont{Boehley}},
\bibinfo{author}{\bibfnamefont{W. J.} \bibnamefont{Amstrong}},
  \bibinfo{journal}{Rev. Sci. Inst.} \textbf{\bibinfo{volume}{75}},
  \bibinfo{number}{11} \bibinfo{pages}{4916}
  (\bibinfo{year}{2004})
  
   \bibitem[{\citenamefont{Miller}(2007)}]{Miller07}
\bibinfo{author}{\bibfnamefont{J. E.}~\bibnamefont{Miller}},
\bibinfo{author}{\bibfnamefont{T. R.} \bibnamefont{Boehly}},
\bibinfo{author}{\bibfnamefont{A.} \bibnamefont{Melchior}},
\bibinfo{author}{\bibfnamefont{D. D.} \bibnamefont{Meyerhofer}},
\bibinfo{author}{\bibfnamefont{P. M.} \bibnamefont{Celliers}},
\bibinfo{author}{\bibfnamefont{J. H.} \bibnamefont{Eggert}},
\bibinfo{author}{\bibfnamefont{D. G.} \bibnamefont{Hicks}},
\bibinfo{author}{\bibfnamefont{C. M.} \bibnamefont{Sorce}},
\bibinfo{author}{\bibfnamefont{J. A.} \bibnamefont{Oertel}},
\bibinfo{author}{\bibfnamefont{P. M.} \bibnamefont{Emmel}},
  \bibinfo{journal}{Rev. Sci. Ins.} ,\textbf{\bibinfo{volume}{78}},
 \bibinfo{pages}{034903} (\bibinfo{year}{2007}).
 
\bibitem{Millot2015a} M. Millot, N. Dubrovinskaia, A. Cernok, S. Blaha, L. Dubrovinsky, D. G. Braun, P. M. Celliers, G. W. Collins, J. H. Eggert, R. Jeanloz, Science, {\bf 347}, 418 (2015).


\bibitem[{\citenamefont{reason}()}]{reason}
\bibinfo{info}{Two effects contribute to modify the SOP throughput on each laser shot involving diamond anvil cell: absorption by the diamond anvil and vignetting. Radiations generated by the ablation plasma can cause photo-ionization of the diamond anvil facing the VISAR/SOP collection optics and attenuate the transmitted thermal emission from the shocked reference and sample. Similarly, optimum geometry of the DAC assembly for mechanical performances (in order to maximize the precompression) can compromise the optical perfomances and introduce vignetting. Measurements relative to the quartz in-situ standard upon breakout of the shock from the quartz to the sample are not affected by these issues.}

   \bibitem[{\citenamefont{Knudson}(2013)}]{KnudsonRelease}
\bibinfo{author}{\bibfnamefont{M. D.} \bibnamefont{Knudson}},
\bibinfo{author}{\bibfnamefont{M. P.} \bibnamefont{Desjarlais}},
  \bibinfo{journal}{Phys. Rev. B} \textbf{\bibinfo{volume}{88}},
 \bibinfo{pages}{184107}
  (\bibinfo{year}{2013}) 
  
 \bibitem{Qi2015} T. Qi, M. Millot, R. G. Krauss, S. Root, S. Hamel, Phys. Plasmas,\textbf{22},062706 (2015)
  
\bibitem[{\citenamefont{Barker}(1965)}]{Barker65}
\bibinfo{author}{\bibfnamefont{L. M.} \bibnamefont{Barker}},
  \bibinfo{journal}{Rev. Sci. Inst.} \textbf{\bibinfo{volume}{36}},
  \bibinfo{number}{11} \bibinfo{pages}{1617}
  (\bibinfo{year}{1965})


\bibitem[{\citenamefont{Barker}(1970)}]{Barker70}
\bibinfo{author}{\bibfnamefont{L. M.} \bibnamefont{Barker}},
\bibinfo{author}{\bibfnamefont{R. E.} \bibnamefont{Hollenbach}},
  \bibinfo{journal}{J. Appl. Phys.} \textbf{\bibinfo{volume}{41}},
 \bibinfo{pages}{10}
  (\bibinfo{year}{1970})

 \bibitem[{\citenamefont{Barker}(1972)}]{Barker72}
\bibinfo{author}{\bibfnamefont{L. M.} \bibnamefont{Barker}},
\bibinfo{author}{\bibfnamefont{R. E.} \bibnamefont{Hollenbach}},
  \bibinfo{journal}{J. Appl. Phys.} \textbf{\bibinfo{volume}{43}},
 \bibinfo{pages}{11}
  (\bibinfo{year}{1972})

 
  
\bibitem[{\citenamefont{Hicks}(2005)}]{Hicks05}
\bibinfo{author}{\bibfnamefont{D. G.} \bibnamefont{Hicks}},
\bibinfo{author}{\bibfnamefont{T. R.} \bibnamefont{Boehly}},
\bibinfo{author}{\bibfnamefont{P. M.} \bibnamefont{Celliers}},
\bibinfo{author}{\bibfnamefont{J. H.} \bibnamefont{Eggert}},
\bibinfo{author}{\bibfnamefont{E.} \bibnamefont{Vianello}},
\bibinfo{author}{\bibfnamefont{D. D.} \bibnamefont{Meyerhofer}},
\bibinfo{author}{\bibfnamefont{G. W.} \bibnamefont{Collins}},
  \bibinfo{journal}{Phys. Plas.} \textbf{\bibinfo{volume}{12}},
 \bibinfo{pages}{082702}
  (\bibinfo{year}{2005})
 
\bibitem[{\citenamefont{Knudson}(2009)}]{Knudson}
\bibinfo{author}{\bibfnamefont{M. D.} \bibnamefont{Knudson}},
\bibinfo{author}{\bibfnamefont{M. P.} \bibnamefont{Desjarlais}},
  \bibinfo{journal}{Phys. Rev. Let.} \textbf{\bibinfo{volume}{103}},
 \bibinfo{pages}{225501}
  (\bibinfo{year}{2009})


\bibitem[{\citenamefont{Adadurov}(1962)}]{Adadurov62}
\bibinfo{author}{\bibfnamefont{G. A.} \bibnamefont{Adadurov}},
\bibinfo{author}{\bibfnamefont{A. N.} \bibnamefont{Dremin}},
\bibinfo{author}{\bibfnamefont{S. V.} \bibnamefont{Pershin}},
\bibinfo{author}{\bibfnamefont{V. N.} \bibnamefont{Rodionov}},
\bibinfo{author}{\bibfnamefont{Yu. N.} \bibnamefont{Ryabinin}},
  \bibinfo{journal}{Zh. Prikl. Mekh. Tekhn. Fiz. } \textbf{\bibinfo{volume}{4}},
 \bibinfo{pages}{81}
  (\bibinfo{year}{1962})

\bibitem[{\citenamefont{Altshuler}(1965)}]{Altshuler65}
\bibinfo{author}{\bibfnamefont{L. V.} \bibnamefont{Al'tshuler}},
  \bibinfo{author}{\bibfnamefont{R. F.}~\bibnamefont{Trunin}},
   \bibinfo{author}{\bibfnamefont{G. V.}~\bibnamefont{Simakov}},
  \bibinfo{journal}{Izv. Akad. Nauk SSSR. Fiz. Zemli } \textbf{\bibinfo{volume}{10}},
  \bibinfo{pages}{1} (\bibinfo{year}{1965}).

\bibitem[{\citenamefont{Trunin}(1971)}]{Trunin71}
  \bibinfo{author}{\bibfnamefont{R. F.}~\bibnamefont{Trunin}},
   \bibinfo{author}{\bibfnamefont{G. V.}~\bibnamefont{Simakov}},
   \bibinfo{author}{\bibfnamefont{M. A.} \bibnamefont{Produrets}},
 \bibinfo{author}{\bibfnamefont{B. N.} \bibnamefont{Moiseev}},
 \bibinfo{author}{\bibfnamefont{L. V.} \bibnamefont{Popov}},
  \bibinfo{journal}{Izv. Akad. Nauk SSSR. Fiz Zemli} \textbf{\bibinfo{volume}{1}},
  \bibinfo{pages}{13} (\bibinfo{year}{1971}).

 \bibitem[{\citenamefont{Pavlovskii}(1976)}]{Pavlovskii76}
  \bibinfo{author}{\bibfnamefont{M. N.}~\bibnamefont{Pavlovskii}},
  \bibinfo{journal}{Zh. Prikl. Mekh. Tekhn. Fiz. } \textbf{\bibinfo{volume}{5}},
  \bibinfo{pages}{136} (\bibinfo{year}{1971}).

 \bibitem[{\citenamefont{VanThiel}(1977)}]{VanThiel77}
  \bibinfo{author}{\bibfnamefont{M. }~\bibnamefont{Van Thiel}},
  \bibinfo{journal}{LLNL Report UCRL } \textbf{\bibinfo{volume}{50108}},
  \bibinfo{pages}{373} (\bibinfo{year}{1977}).


 \bibitem[{\citenamefont{Marsh}(1980)}]{Marsh80}
\bibinfo{author}{\bibfnamefont{S. P.}~\bibnamefont{Marsh}},
  \emph{\bibinfo{title}{LASL Shock Hugoniot Data}}
  (\bibinfo{publisher}{Univ. California Press, Berkeley}, \bibinfo{year}{1980}).

\bibitem[{\citenamefont{Trunin}(1994)}]{Trunin94}
  \bibinfo{author}{\bibfnamefont{R. F.}~\bibnamefont{Trunin}},
  \bibinfo{journal}{Usp. Fiz. Nauk } \textbf{\bibinfo{volume}{164}},
  \bibinfo{number}{11},  \bibinfo{pages}{1215} (\bibinfo{year}{1994}).


\bibitem[{\citenamefont{Kerley}(1999)}]{Kerley99}
\bibinfo{author}{\bibfnamefont{G.}~\bibnamefont{Kerley}} ,
  \emph{\bibinfo{title}{Equations of state for composite Materials }}
  (\bibinfo{publisher}{Kerley, Albuquerque, NM}, \bibinfo{year}{1999}).
  



\bibitem[{\citenamefont{Trunin}(1970)}]{Trunin70}
  \bibinfo{author}{\bibfnamefont{R. F.}~\bibnamefont{Trunin}},
   \bibinfo{author}{\bibfnamefont{G. V.}~\bibnamefont{Simakov}},
   \bibinfo{author}{\bibfnamefont{M. A.} \bibnamefont{Produrets}},
 \bibinfo{author}{\bibfnamefont{B. N.} \bibnamefont{Moiseev}},
 \bibinfo{author}{\bibfnamefont{L. V.} \bibnamefont{Popov}},
  \bibinfo{journal}{Earth Physics} ,
  \bibinfo{pages}{1} (\bibinfo{year}{1970}).

\bibitem[{\citenamefont{Trunin}(1994)}]{Trunin94b}
  \bibinfo{author}{\bibfnamefont{R. F.}~\bibnamefont{Trunin}},
  \bibinfo{journal}{Physics-Uspekhi} ,\textbf{\bibinfo{volume}{37}},
     \bibinfo{number}{11} ,\bibinfo{pages}{1123} (\bibinfo{year}{1994}).


\bibitem[{\citenamefont{Luo}(2002)}]{Luo02}
\bibinfo{author}{\bibfnamefont{S-N.} \bibnamefont{Luo}},
\bibinfo{author}{\bibfnamefont{J.} \bibnamefont{Mosenfelder}},
\bibinfo{author}{\bibfnamefont{P.} \bibnamefont{Asimov}},
\bibinfo{author}{\bibfnamefont{T.} \bibnamefont{Ahrens}},
  \bibinfo{journal}{Physics-Uspekhi} \textbf{\bibinfo{volume}{45}},
  \bibinfo{pages}{435}, (\bibinfo{year}{2002}).


\bibitem[{\citenamefont{Luo}(2002)}]{Luo02b}
\bibinfo{author}{\bibfnamefont{S-N.} \bibnamefont{Luo}},
\bibinfo{author}{\bibfnamefont{J.} \bibnamefont{Mosenfelder}},
\bibinfo{author}{\bibfnamefont{P.} \bibnamefont{Asimov}},
\bibinfo{author}{\bibfnamefont{T.} \bibnamefont{Ahrens}},
  \bibinfo{journal}{Geophysical research letters} \textbf{\bibinfo{volume}{29}},
   \bibinfo{number}{14} , \bibinfo{pages}{1691} (\bibinfo{year}{2002}).



  \bibitem[{\citenamefont{gruneisen}()}]{gruneisen}
\bibinfo{author}{\bibfnamefont{E.}~\bibnamefont{Gruneisen}},
  \emph{\bibinfo{title}{Handbuch der Physik}}
  (\bibinfo{publisher}{ed. H Greiger and K . Scheel, Springer}, \bibinfo{year}{1980}).
\bibinfo{number}{10},  \bibinfo{pages}{1}

\bibitem{Barrios2010} M. A. Barrios, D. Hicks, T. R. Boehly, D. E. Fratanduono, J. H. Eggert, P. M. Celliers, G. W. Collins, and D. D.Meyerhofer, Phys. Plas., {\bf 17}, 056307 (2010).


  \bibitem[{\citenamefont{Hicks}(2006)}]{Hicks06}
\bibinfo{author}{\bibfnamefont{D. G.} \bibnamefont{Hicks}},
\bibinfo{author}{\bibfnamefont{T. R.} \bibnamefont{Boehly}},
\bibinfo{author}{\bibfnamefont{J. H.} \bibnamefont{Eggert}},
\bibinfo{author}{\bibfnamefont{J. E.} \bibnamefont{Miller}},
\bibinfo{author}{\bibfnamefont{P. M.} \bibnamefont{Celliers}},
\bibinfo{author}{\bibfnamefont{G. W.} \bibnamefont{Collins}},
  \bibinfo{journal}{Phys. Rev. Let.} \textbf{\bibinfo{volume}{97}},
 \bibinfo{pages}{025502}
  (\bibinfo{year}{2006})
 
  \bibitem{Militzer} B. Militzer, S. Mazevet and P. Loubeyre, Phys. Rev. B {\bf 79}, 155105 (2009).
 
 \bibitem{SCVH} D. Saumon, G. Chabrier and H. Van Horn, Astrophys. J. {\bf 99}, 713 (1995). 
 
 \bibitem{Caillabet} L. Caillabet, S. Mazevet and P. Loubeyre, Phys. Rev. B {\bf 83}, 094101 (2011).
 
  \bibitem[{\citenamefont{}()}]{TempGrun}
     From the thermodynamic definition of $\Gamma$,
     $\Gamma=V\left(\frac{\partial P}{\partial E}\right)_{V}$ or $\left(\frac{\partial P}{\partial T}\right)_{V} = \frac{\Gamma
     C_{V}}{V}$ and the expression $dT=\left(\frac{\partial T}{\partial
     V}\right)_{S}dV+\left(\frac{\partial T}{\partial
     S}\right)_{V}dS$, we can show that along the Hugoniot $\frac{dT_{H}}{dV}+\frac{\Gamma T_{H}}{V}+\frac{q_{H}}{C_{V}}=0$ with $q_{H}=-\frac{ S_{H}}{V}=-P_{H}-\frac{dE_{H}}{dV}$


  
\bibitem [{\citenamefont {Gregoryanz}\ \emph {et~al.}(2000)\citenamefont
  {Gregoryanz}, \citenamefont {Hemley}, \citenamefont {Mao},\ and\
  \citenamefont {Gillet}}]{Gregoryanz2000a}%
  \BibitemOpen
  \bibfield  {author} {\bibinfo {author} {\bibfnamefont {E.}~\bibnamefont
  {Gregoryanz}}, \bibinfo {author} {\bibfnamefont {R.~J.}\ \bibnamefont
  {Hemley}}, \bibinfo {author} {\bibfnamefont {H.-k.}\ \bibnamefont {Mao}}, \
  and\ \bibinfo {author} {\bibfnamefont {P.}~\bibnamefont {Gillet}},\ }\href
  {http://www.ncbi.nlm.nih.gov/pubmed/12633276} {\bibfield  {journal} {\bibinfo
   {journal} {Phys. Rev. Let.}\ }\textbf {\bibinfo {volume} {84}},\
  \bibinfo {pages} {3117} (\bibinfo {year} {2000})}
  
\bibitem [{\citenamefont {Bottom}(1982)}]{Bottom1982}%
  \BibitemOpen
  \bibfield  {author} {\bibinfo {author} {\bibfnamefont {V.~E.}\ \bibnamefont
  {Bottom}},\ } {\emph {\bibinfo {title} {{Introduction to quartz
  crystal unit design}}}}\ (\bibinfo  {publisher} {Van Nostrand},\ \bibinfo
  {address} {New York, NY},\ \bibinfo {year} {1982})\ p.\ \bibinfo {pages}
  {265}
  
\bibitem [{\citenamefont {Barron}\ \emph {et~al.}(1982)\citenamefont {Barron},
  \citenamefont {Collins}, \citenamefont {W},\ and\ \citenamefont
  {K}}]{Barron1982}%
  \BibitemOpen
  \bibfield  {author} {\bibinfo {author} {\bibfnamefont {T.~H.~K.}\
  \bibnamefont {Barron}}, \bibinfo {author} {\bibfnamefont {J.~F.}\
  \bibnamefont {Collins}}, \bibinfo {author} {\bibfnamefont {S.~T.}\
  \bibnamefont {W}}, \ and\ \bibinfo {author} {\bibfnamefont {W.~G.}\
  \bibnamefont {K}},\ }\href {\doibase doi:10.1088/0022-3719/15/20/016}
  {\bibfield  {journal} {\bibinfo  {journal} {J. Phys. C: Solid State Phys.}\
  }\textbf {\bibinfo {volume} {15}},\ \bibinfo {pages} {4311} (\bibinfo {year}
  {1982})}
  
\bibitem [{\citenamefont {Calderon}\ \emph {et~al.}(2007)\citenamefont
  {Calderon}, \citenamefont {Gauthier}, \citenamefont {Decremps}, \citenamefont
  {Hamel}, \citenamefont {Syfosse},\ and\ \citenamefont
  {Polian}}]{Calderon2007}%
  \BibitemOpen
  \bibfield  {author} {\bibinfo {author} {\bibfnamefont {E.}~\bibnamefont
  {Calderon}}, \bibinfo {author} {\bibfnamefont {M.}~\bibnamefont {Gauthier}},
  \bibinfo {author} {\bibfnamefont {F.}~\bibnamefont {Decremps}}, \bibinfo
  {author} {\bibfnamefont {G.}~\bibnamefont {Hamel}}, \bibinfo {author}
  {\bibfnamefont {G.}~\bibnamefont {Syfosse}}, \ and\ \bibinfo {author}
  {\bibfnamefont {A.}~\bibnamefont {Polian}},\ } {\bibfield
  {journal} {\bibinfo  {journal} {Journal of Physics: Condensed Matter}\
  }\textbf {\bibinfo {volume} {19}},\ \bibinfo {pages} {436228} (\bibinfo
  {year} {2007})}
  
\bibitem [{\citenamefont {Kimizuka}\ \emph {et~al.}(2007)\citenamefont
  {Kimizuka}, \citenamefont {Ogata}, \citenamefont {Li},\ and\ \citenamefont
  {Shibutani}}]{Kimizuka2007}%
  \BibitemOpen
  \bibfield  {author} {\bibinfo {author} {\bibfnamefont {H.}~\bibnamefont
  {Kimizuka}}, \bibinfo {author} {\bibfnamefont {S.}~\bibnamefont {Ogata}},
  \bibinfo {author} {\bibfnamefont {J.}~\bibnamefont {Li}}, \ and\ \bibinfo
  {author} {\bibfnamefont {Y.}~\bibnamefont {Shibutani}},\ }\href {\doibase
  10.1103/PhysRevB.75.054109} {\bibfield  {journal} {\bibinfo  {journal}
  {Phys. Rev. B}\ }\textbf {\bibinfo {volume} {75}},\ \bibinfo {pages} {1}
  (\bibinfo {year} {2007})}
  
\bibitem [{\citenamefont {Ghosh}(1999)}]{Ghosh1999}%
  \BibitemOpen
  \bibfield  {author} {\bibinfo {author} {\bibfnamefont {G.}~\bibnamefont
  {Ghosh}},\ }\href {\doibase 10.1016/S0030-4018(99)00091-7} {\bibfield
  {journal} {\bibinfo  {journal} {Optics Communications}\ }\textbf {\bibinfo
  {volume} {163}},\ \bibinfo {pages} {95} (\bibinfo {year} {1999})}
  
\bibitem [{\citenamefont {Vedam}\ and\ \citenamefont
  {Davis}(1967)}]{Vedam1967}%
  \BibitemOpen
  \bibfield  {author} {\bibinfo {author} {\bibfnamefont {K.}~\bibnamefont
  {Vedam}}\ and\ \bibinfo {author} {\bibfnamefont {T. A.}\ \bibnamefont
  {Davis}},\ }\href {\doibase 10.1364/JOSA.57.001140} {\bibfield  {journal}
  {\bibinfo  {journal} {Journal of the Optical Society of America}\ }\textbf
  {\bibinfo {volume} {57}},\ \bibinfo {pages} {1140} (\bibinfo {year}
  {1967})}
  
\bibitem[{\citenamefont{noteIndex}()}]{noteIndex}
\BibitemOpen
{We recalculate the quartz refractive index pressure dependence from interferometry data up to 0.7 GPa\cite{Vedam1967}. We use the most recent elastic constant \cite{Calderon2007} instead of Bridgman's values\cite{Bridgman1928} to infer the quartz plate thickness and the density under pressure. We find a $1.8\%$ lower linear compressibility at max pressure that translates into a slightly slower increase of the index with pressure. We assume that the optical dispersion between 589 nm (wavelength at which the pressure dependence has been measured) and 532 nm is pressure independent in the range explored here ($p<10$ GPa). A simple linear fit of the index as a function of density\cite{Dewaele2003} is then used to extrapolate up to the initial density $\rho_0$.}
		
  

  
\bibitem [{\citenamefont {Vedam}\ and\ \citenamefont
  {Samara}(1983)}]{Vedam1983}%
  \BibitemOpen
  \bibfield  {author} {\bibinfo {author} {\bibfnamefont {K.}~\bibnamefont
  {Vedam}}\ and\ \bibinfo {author} {\bibfnamefont {G.~A.}\ \bibnamefont
  {Samara}},\ } {\bibfield  {journal} {\bibinfo  {journal}
  {Critical Reviews in Solid State and Materials Sciences}\ }\textbf {\bibinfo
  {volume} {11}},\ \bibinfo {pages} {1} (\bibinfo {year} {1983})}
  

\bibitem [{\citenamefont {Bridgman}(1928)}]{Bridgman1928}%
  \BibitemOpen
  \bibfield  {author} {\bibinfo {author} {\bibfnamefont {P.~W.}\ \bibnamefont
  {Bridgman}},\ } {\bibfield  {journal} {\bibinfo  {journal}
  {American journal of science}\ }\textbf {\bibinfo {volume} {15}},\ \bibinfo
  {pages} {287} (\bibinfo {year} {1928})}
  
\bibitem [{\citenamefont {Dewaele}\ \emph {et~al.}(2003)\citenamefont
  {Dewaele}, \citenamefont {Eggert}, \citenamefont {Loubeyre},\ and\
  \citenamefont {{Le Toullec}}}]{Dewaele2003}%
  \BibitemOpen
  \bibfield  {author} {\bibinfo {author} {\bibfnamefont {A.}~\bibnamefont
  {Dewaele}}, \bibinfo {author} {\bibfnamefont {J.}~\bibnamefont {Eggert}},
  \bibinfo {author} {\bibfnamefont {P.}~\bibnamefont {Loubeyre}}, \ and\
  \bibinfo {author} {\bibfnamefont {R.}~\bibnamefont {{Le Toullec}}},\ }\href
  {\doibase 10.1103/PhysRevB.67.094112} {\bibfield  {journal} {\bibinfo
  {journal} {Phys. Rev. B}\ }\textbf {\bibinfo {volume} {67}},\ \bibinfo
  {pages} {1} (\bibinfo {year} {2003})}


  
\end{thebibliography}
\end{document}